\documentclass[twocolumn]{aastex631}
\newcommand{\sftw}[1]{\texttt{#1}}

\usepackage{hyperref}
\usepackage{color}
\usepackage[normalem]{ulem}
\usepackage{amsmath,amssymb}	
\usepackage{mathpazo}
\usepackage{textcomp}
\usepackage{booktabs}
\usepackage{xcolor}
\usepackage{nicefrac}
\usepackage{bm}
\usepackage{graphicx}
\usepackage{url}
\usepackage{savesym}
\usepackage[T1]{fontenc}

\definecolor{green1}{RGB}{0, 128, 0}

\shorttitle{No self-shadowing instability in 2D models of irradiated protoplanetary disks}
\shortauthors{Melon Fuksman, Klahr}

\begin{document}

%\title{Radiation-hydrodinamical instabilities in circumstellar disk atmospheres\\I. Self-shadowing instability}
\title{No self-shadowing instability in 2D radiation-hydrodynamical models of irradiated protoplanetary disks}

\correspondingauthor{Julio David Melon Fuksman}
\email{fuksman@mpia.de}

\author[0000-0002-1697-6433]{Julio David Melon Fuksman}
\affil{Max Planck Institute for Astronomy, K\"onigstuhl 17, 69117 Heidelberg, Germany}

\author[0000-0002-8227-5467]{Hubert Klahr}
\affil{Max Planck Institute for Astronomy, K\"onigstuhl 17, 69117 Heidelberg, Germany}

\begin{abstract}
Theoretical models of protoplanetary disks including stellar irradiation often show a spontaneous amplification of scale height perturbations, produced by the enhanced absorption of starlight in enlarged regions. In turn, such regions cast shadows on adjacent zones that consequently cool down and shrink, eventually leading to an alternating pattern of overheated and shadowed regions. Previous investigations have proposed this to be a real self-sustained process, the so-called self-shadowing or thermal wave instability, which could naturally form frequently observed disk structures such as rings and gaps, and even potentially enhance the formation of planetesimals. All of these, however, have assumed in one way or another vertical hydrostatic equilibrium and instantaneous radiative diffusion throughout the disk. In this work we present the first study of the stability of accretion disks to self-shadowing that relaxes these assumptions, relying instead on radiation-hydrodynamical simulations. We first construct hydrostatic disk configurations by means of an iterative procedure and show that the formation of a pattern of enlarged and shadowed regions is a direct consequence of assuming instantaneous radiative diffusion. We then let these solutions evolve in time, which leads to a fast damping of the initial shadowing features in layers close to the disk surface. These thermally relaxed layers grow towards the midplane until all temperature extrema in the radial direction are erased in the entire disk. Our results suggest that radiative cooling and gas advection at the disk surface prevent a self-shadowing instability from forming, by damping temperature perturbations before these reach lower, optically thick regions.

\end{abstract}


 \section{Introduction}\label{S:Introduction}

Understanding the different physical processes behind the structure and evolution of protoplanetary disks is of fundamental importance to explain the formation of planets. Gas dynamics in disks around young stars affects the transport and accumulation of dust, thereby shaping the resulting planetary systems \citep[a recent review on hydro-, magnetohydro-, and gas-dust dynamics in protoplanetary disks can be found in][]{Lesur2022}. Given that these disks are largely cold and poorly ionized \citep{Dzyurkevich2013,Desch2015}, it is currently unclear to what extent magnetic phenomena such as the magnetorotational instability (MRI) \citep[][]{Velikhov1959,Chandrasekhar1960,BalbusHawley1991}, the Hall-shear instability \citep[][]{Kunz2008,Lesur2014}, and magnetocentrtifugal winds \citep[see e.g.][]{Blandford1982,Suzuki2009,Bai2013,Fromang2013,Simon2016} can be responsible for the observed accretion rates onto the central star, particularly considering the stabilizing factor of nonideal magnetic effects such as resistivity and ambipolar diffusion \citep{Turner2014,Lesur2014,Simon2015,Lesur2021}. It is therefore likely that the formation of structures and the turbulent transport of angular momentum in large, magnetically inactive "dead" or "Ohmic" zones are largely controlled by hydrodynamical instabilities \citep{Klahr2018,Lyra2019,Pfeil2019}, such as the subcritical baroclinic instability \citep[][]{KlahrBodenheimer2003,Petersen2007I,Petersen2007II,Lesur2010}, the convective overstability (COS) \citep[][]{Klahr2014,Lyra2014}, the zombie vortex instability \citep{Marcus2015,Marcus2016}, the vertical shear instability (VSI) \citep[][]{Urpin1998,Nelson2013,Stoll2014,Flores2020,Manger2020,Manger2021},
%originally studied in the context of rotating stars by \cite{Goldreich1967} and \cite{Fricke1968},
and the recently discovered local VSI (Klahr et al. 2022, submitted) and nonlinear symmetric instability (Klahr \& Baehr 2022, submitted). Besides producing turbulence and limiting dust settling \citep[see, e.g.,][]{Birnstiel2010,Flock2020}, hydrodynamical instabilities may trigger the formation of long-lived structures such as vortices and zonal flows \citep[see, e.g.,][]{Manger2018,Pfeil2021} that due to their pressure structure can accumulate dust and pebbles, potentially accelerating planetesimal formation via gravitational collapse \citep{Dittrich2013,Raettig2015,Gerbig2020,Raettig2021}. Furthermore, a clear understanding of these phenomena is crucial to explain current observations of structures in protoplanetary disks \citep{Andrews2018,Barraza2021} and help confirm or rule out the detection of exoplanets, as it is often unclear if planetary perturbations are required to explain such features \citep{Calcino2020,Pyerin2021,BrownSevilla2021}.

An important, yet unanswered question in this picture is if protoplanetary disks are globally stable to local perturbations in their scale height. Local regions of enlarged aspect ratio absorb stellar radiation at a higher rate than their surroundings, potentially heating up and further inducing a local growth of the disk's width. Conversely, shadowed regions cool down, which causes them to shrink and absorb even less starlight. In all models of this process, perturbations move inwards due to their asymmetric irradiation: any given enlarged region absorbs more starlight on their inner face, i.e., closer to the star, than on their shadowed outer side. As a consequence, the inner side grows and the outer one shrinks, shifting the center of the enlarged region towards the star. It is currently unknown if this feedback process can be efficient enough to trigger a self-sustained instability or if, on the contrary, the formation of scale height perturbations is damped by even faster radiation-hydrodynamical processes. 

This formation and evolution of temperature bumps (or thermal waves) has received different names in literature, such as thermal wave instability \citep{Watanabe2008,Ueda2021}, iradiation instability \citep{WuLitwick2021}, and self-shadowing instability \citep{Dullemond2000}. We shall use the latter as we find it the most self-explanatory. It was argued by several authors that such an instability could have major consequences for planet formation, either by affecting the composition of formed planetesimals by inducing temporal variations in snow line locations \citep{Ueda2021,Okuzumi2022}, boosting the formation of planetesimals by creating dust traps \citep{Okuzumi2022}, or triggering hydrodynamical instabilities in the dead zone \citep{Watanabe2008}. On the other hand, as suggested by \cite{Siebenmorgen2012} and \cite{Ueda2021}, such an instability could naturally produce rings and gaps such as those observed in dust continuum  \citep{Andrews2018} and scattered light \citep{Avenhaus2018} images of protoplanetary disks. 

The stability of circumstellar disks to self-shadowing was first addressed by \cite{DAlessio1999}, who studied temperature perturbations in vertically isothermal disks. Using typical parameters for accreting T-Tauri stars, these authors assumed
instantaneous vertical hydrostatic equilibrium in the inner regions of the disk ($r\lesssim 2$ au), where they estimated the hydrostatic equilibrium timescale $t_\mathrm{dyn}$ to be much shorter than the radiative cooling timescale $t_\mathrm{therm}$. Assuming axisymmetry, they constructed and numerically solved a 1D model for the evolution of the midplane temperature, considering absorption of stellar irradiation
at an altitude $z_s$ above the midplane and radiative cooling. This analysis concluded that temperature perturbations travel inward as they get damped, which suggested that disks are stable under such circumstances.
Shortly after, \cite{Dullemond2000} studied the opposite limit of this model, i.e., $t_\mathrm{dyn}\gg t_\mathrm{therm}$, therefore computing the midplane temperature from a thermal equilibrium condition and constructing a simplified equation for the acceleration of the pressure scale height due to its departure from equlibrium. A linear analysis of that model showed that inward-propagating modes grow exponentially, which supported the instability hypothesis.

A new take on the hydrostatic case ($t_\mathrm{dyn}\ll t_\mathrm{therm}$) was conducted by \cite{Watanabe2008}, who relaxed the assumption made by the previous authors that the $z_s/H$ ratio, where $H$ is the pressure scale height, is constant throughout the disk. Instead, these authors obtained $z_s$ numerically by computing the optical depth $\tau_s$ at the star's temperature along straight lines from the star and defining the irradiation surface as the location of space verifying $\tau_s=1$.
%They employed as well a more refined treatment for the irradiation heating and the radiative losses from the disk by accounting for the vertical dust optical depth at the star and midplane temperatures.
The main results of this model contradicted the conclusions of \cite{DAlessio1999}, as they showed a sustained spontaneous formation of inward-propagating temperature waves within $100$ au from the central star.
%Depending on the introduced viscous heating, such waves could be dampened at small radii and even became stationary for high-enough accretion rates.
As argued by the authors and later supported by the linear analysis by \citep{WuLitwick2021}, this different behavior is directly caused by the variation of the $z_s/H$ ratio with temperature perturbations. 

Later improvements to this 1D model were made by \cite{Ueda2021}, \cite{WuLitwick2021}, and \cite{Okuzumi2022}. The first of these authors considered a quite similar model including external heating from the surrounding cloud, which suppressed the instability at large radii for low-enough dust content (dust-to-gas mass ratio of $\sim 10^{-4}$ in their case). On the other hand, \cite{WuLitwick2021} computed the stellar irradiation heating flux via the Stefan-Boltzmann law with a temperature resulting from 2D Monte Carlo simulations carried out at each time step, and increased the thermal timescale used in their evolution equation to mimic the effect of vertical radiative diffusion. Using density values consistent with observed systems, they obtained temperature waves within $30$ au from the star ($\sim 100$ au if dust settling was included), outside of which the instability was suppressed by radial radiation transport. Finally, \cite{Okuzumi2022} included the same timescale correction for vertical diffusion as in \cite{WuLitwick2021}, and computed stellar heating 
%by approximating the reprocessed starlight flux
by means of an approximate radiative transfer model based on
%. Instead of radially averaging the downward approximate flux over a selected characteristic distance, as done in \cite{Watanabe2008,Ueda2021} using $z_s$, 
an integration of the incident flux from the entire irradiation surface onto each element of the reprocessed radiation photosphere. Their results are overall similar to those obtained by the other authors, showing inward-moving temperature waves up to $30$ au with sharper variations than in previous works due to the different radiative transfer scheme. As suggested by \cite{Ueda2021} and supported by the linear analysis of \cite{WuLitwick2021}, the strength and growth rate of the instability in these models increase for higher $z_s/H$, which is naturally verified for high dust contents and therefore high opacities. Moreover, high opacities reduce radial radiation transport, which otherwise damps the formation of temperature bumps.

Similar unstable behavior is observed in the 2D hydrostatic models of protoplanetary disks by \cite{DullemondDominik2004,Min2009,Siebenmorgen2012,Ueda2019}, with the difference that these are not time-dependent. Instead, these authors produced hydrostatic configurations by iteratively alternating the computation of temperature distributions via Monte Carlo radiative transfer and the solution of hydrostatic equilibrium equations. As shown in these works, this procedure does not necessarily converge, as the resulting solutions exhibit bumps in the scale height that get shifted towards the star after each iteration. This phenomenon becomes more important for increasing disk masses \citep{Min2009} and is more prominent in T Tauri stars than in Herbig Ae/Be stars \citep{DullemondDominik2004,Siebenmorgen2012}.
Contrarily to the suggestion made by these authors that such temperature oscillations are related to the self-shadowing instability, \cite{WuLitwick2021} argue that these may also be influenced by an intrinsic numerical instability caused by the iterative procedure, and that disks that are unstable to self-shadowing in iterative models may be stable in reality. Similar shadowing features can also be seen in the Monte Carlo-based iterative method applied in \cite{Isella2018}, in which the 3D gas distribution is reconstructed at each step from 2D hydrodynamics simulations of the midplane density assuming vertical hydrostatic equilibrium.

Several effects disregarded in all of these works could potentially prevent the formation of the self-shadowing instability. The assumption of a vertically hydrostatic disk has been so far justified on timescales estimations, but none of these studies involved the solution of 3D (or 2D axisymmetric) hydrodynamics equations. Therefore, meridional advection of mass and energy, which could in principle kill off any large-scale perturbation via sound waves, has so far been disregarded. On the other hand, all models described above assume in one way or another instant information, meaning that the midplane temperature is at any point in time immediately affected by the temperature at the photosphere or even at distant radii. This is at odds with the fact that energy is transported via radiative transfer and advection over finite times. It is then unclear if surface temperature perturbations reach the midplane fast enough to cause a fast, substantial growth in the local scale height before they change at the surface. Thus, hydrostatic models may enhance self-shadowing effects by neglecting such delay. On this regard, a linear analysis made by \cite{WuLitwick2021} shows that vertical temperature waves due to surface perturbations are damped on their way to the midplane, although this study does not conclusively answer if that is enough to suppress the instability. It appears clear that answering if self-shadowing can lead to a physical instability requires a simultaneous treatment of stellar irradiation, radiative transfer of the disk's own thermal emission, and hydrodynamics. 

In this work we address this issue by solving the equations of radiation hydrodynamics (Rad-HD) with stellar irradiation. Similar simulations in 3D and in optically thin disk regions ($>20$ au), where self-shadowing is not expected, have been made by \cite{Flock2017VSI}, obtaining VSI-induced turbulence. We assume axisymmetry and study the disk's stability in 2D simulations of the region between $0.4$ and $100$ au. In this way, we do not attempt to make a self-consistent description of the disk's inner rim, which we assume to be much closer to the central star. We note that 2D axisymmetric Rad-HD simulations of the inner rim have been made in \cite{Flock2016InnerRim} including temperature-dependent viscosity, obtaining steady state configurations that remain stable for thousands of dynamical timescales. Further 3D studies of the inner rim employing nonideal radiation magnetohydrodynamics without viscosity \citep{Flock2017InnerRimMHD} obtained steady-state MRI-induced turbulence and vortex formation via the Rossby wave instability. In this work we neglect magnetic fields in order to focus on instability solely due to shadowing effects, and we include no viscosity to reduce damping factors as much as possible. We employ for this the two-moment Rad-HD scheme described in \cite{MelonFuksman2021}, which is able to maintain the direction of freely streaming radiation flows.
This avoids the artificial spreading of radiation energy produced by simpler moment-based radiative transfer schemes such as flux-limited diffusion \citep{Hayes2003}, which could in principle enhance the damping of thermal waves.

This paper is organized as follows. In Section \ref{S:RadHD}, we summarize the equations of our model and the employed numerical methods. In Section \ref{S:DiskModels}, we construct hydrostatic models of protoplanetary disks and characterize the development of features produced by self-shadowing. 
In Section \ref{S:ResultsRadHD} we run Rad-HD simulations to study the time evolution of our hydrostatic configurations and discuss the implications and limitations of our results in the context of the disk's stability to self-shadowing. In Section \ref{S:Conclusions}, we summarize the main results of our work. Additional implementation details and tests at increased resolution are included in the Appendix.
 \section{Radiation hydrodynamics}\label{S:RadHD}
 
 We model the gas in the circumstellar disk and the absorption and emission
 of radiation by means of the nonrelativistic Rad-HD
 module \citep{MelonFuksman2021} implemented in the PLUTO code \citep{Mignone2007}. In the configuration used in this work, the Rad-HD equations read:
 \begin{equation}\label{Eq:RadHD}
 \begin{split}
\frac{\partial \rho}{\partial t} + \nabla \cdot 
\left(\rho \mathbf{v}\right) &= 0 \\
\frac{\partial( \rho \mathbf{v})}{\partial t} + \nabla \cdot 
\left(\rho \mathbf{v} \mathbf{v}\right)+\nabla p &= 
   \mathbf{G}     -\rho\nabla \Phi  \\
\frac{\partial\left( E+\rho\Phi\right)}{\partial t} + \nabla \cdot 
\left[(E+p+\rho\Phi) \mathbf{v}\right] &= c\,G^0  
  -\nabla\cdot\mathbf{F}_\mathrm{Irr} \\
\frac{1}{\hat{c}}\frac{\partial E_r}{\partial t}+\nabla\cdot \mathbf{F}_r &= -G^0 \\
\frac{1}{\hat{c}}\frac{\partial \mathbf{F}_r}{\partial t}+\nabla\cdot \mathbb{P}_r &= -\mathbf{G}\,,
\end{split}
\end{equation}
where $\rho$, $p$, and $\mathbf{v}$ are the gas density, pressure and velocity, while $E_r$, $\mathbf{F}_r$, and $\mathbb{P}_r$ are respectively the radiation energy, flux, and pressure tensor.
%The $\Phi$ and $\mathbf{F}_\mathrm{Irr}$ fields represent, respectively, the gravitational potential and irradiation flux of the star.
The gas energy is computed as
\begin{equation}\label{Eq:GasEnergyDensity}
E = \rho \epsilon + \frac{1}{2}\rho \mathbf{v}^2\,,
\end{equation}
where $\rho\epsilon$ is the gas internal energy density.
This quantity is determined as a function of the pressure
by applying the ideal gas equation of state
\begin{equation}\label{Eq:EoS}
    \rho \epsilon = \frac{p}{\Gamma - 1}\,,
\end{equation}
where we take the specific heat ratio $\Gamma$ to be
constant. The radiation pressure
tensor is computed as a function of $E_r$ and $\mathbf{F}_r$
by means of the M1 closure \citep{Levermore1984M1},
as
\begin{equation}\label{Eq:M11}
  P_r^{ij}=D^{ij}E_r\,,
\end{equation}
where the Eddington tensor $D^{ij}$ is defined as
\begin{equation}
D^{ij}=\frac{1-\xi}{2}\,\delta^{ij}+
\frac{3\xi-1}{2}n^in^j\,,
\end{equation}
with
\begin{equation}\label{Eq:M13}
\xi=\frac{3+4f^2}{5+2\sqrt{4-3f^2}}\,,
\end{equation}
where $\bm{n}=\mathbf{F}_r/\vert\vert\mathbf{F}_r\vert\vert$,
$f=\vert\vert\mathbf{F}_r\vert\vert/E_r$, and
$\delta^{ij}$ is the Kronecker delta. We have normalized the radiation flux by $c$ in such a way that both $E_r$ and $\mathbf{F}_r$ are measured in energy density units.

The quantities $c$ and $\hat{c}$ in  Eq. \eqref{Eq:RadHD} correspond respectively
to the real and reduced values of the speed of light. These are chosen in such a way that $\hat{c}/c<1$ in order to reduce the scale disparity between gas and radiation characteristic speeds, thus avoiding the employment of far too small time steps to achieve reasonable run times \citep[see details in][]{MelonFuksman2021}.

We assume axial symmetry around the disk polar axis and solve the Rad-HD equations in 2D using spherical coordinates $(r,\theta)$, where $r$ is the spherical radius and $\theta$ is the polar angle. However, we compute the evolution of all three components of the vectorial fields $\mathbf{v}$ and $\mathbf{F}_r$, i.e., including the azimuthal components ($\phi$-direction). We define in these coordinates the gravitational potential of a star centered at $r=0$ as
\begin{equation}
  \Phi = -\frac{M_s G}{r}\,,
\end{equation}
where $M_s$ and $G$ are respectively the stellar mass and the
gravitational constant.

\begin{figure}[t!]
\centering
\includegraphics[width=\linewidth]{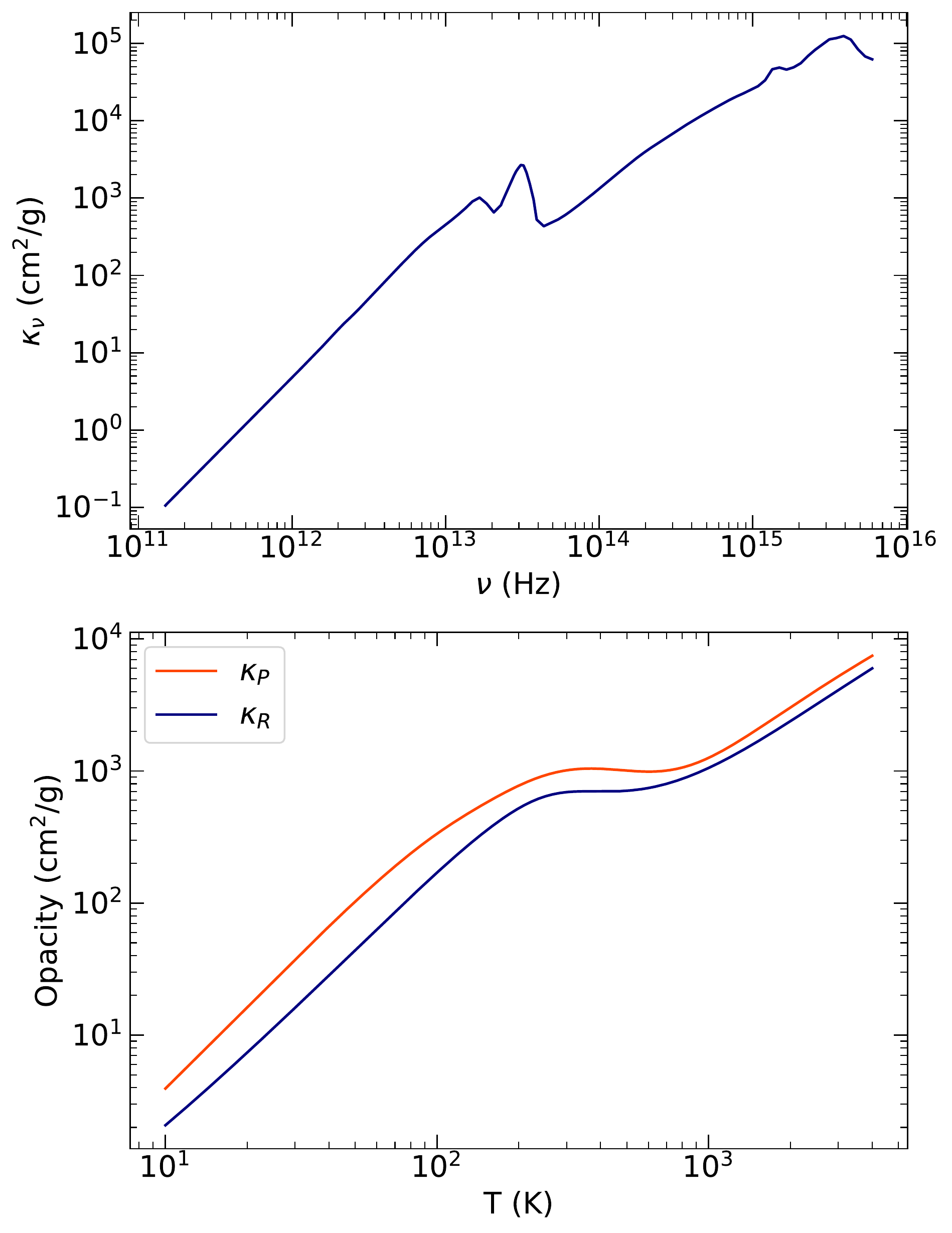}
\caption{Top: Frequency-dependent absorption opacity per dust mass
employed for stellar irradiation.
Bottom: Planck- and Rosseland-averaged 
absorption opacities per dust mass
used to compute radiation-matter interaction terms
(Eq. \eqref{Eq:Gcomov}).}
\label{fig:opacities}
\end{figure}

We include absorption of stellar irradiation by small dust particles modeling the star as a point source and computing the irradiation flux as
\begin{equation}\label{Eq:Firrad}
    \mathbf{F}_\mathrm{Irr}(r,\theta) =
    \pi\left(\frac{R_s}{r}\right)^2
    \int_{\nu_\mathrm{min}}^{\nu_\mathrm{max}} \mathrm{d}\nu\,
    B_\nu(T_s)\,
    e^{-\tau(r,\theta,\nu)}\, \hat{\mathbf{r}} \,,
\end{equation}
where $T_s$ is the star temperature, $R_s$ is the star radius, 
$B_\nu(T)=(2 h \nu^3/c^2)/(e^{h\nu/k_B T}-1)$
is the Planck radiative intensity at a temperature $T$ and frequency $\nu$, and $h$ and $k_B$ are the Planck and Boltzmann constants, respectively.
The frequency-dependent optical depth is computed along radial
trajectories as
\begin{equation}\label{Eq:OpticalDepth}
    \tau(r,\theta,\nu) = 
    \tau_0(r_\mathrm{min},\theta,\nu)+
    \int^r_{r_\mathrm{min}}\mathrm{d}r'\,
    \kappa_\nu\,\rho_\mathrm{d}(r',\theta)\,,
\end{equation}
where $\rho_d$ and $\kappa_\nu$ are respectively the dust density and the frequency-dependent absorption opacity,
while $r_\mathrm{min}$ is the minimum radius of the computational domain.
Since we do not self-consistently model the disk's inner rim, we include in this computation a factor $\tau_0(r_\mathrm{min},\theta,\nu)$ corresponding to the optical depth of the disk portion that is left out of the domain (see Appendix \ref{A:OptDepth}). 
For both the absorption of stellar irradiation at optical and near-infrared wavelengths
and the transfer of reprocessed infrared radiation, dust opacity is dominated by small ($< 1$ $\mu$m) grains. We assume that these are small enough to neglect vertical settling (we discuss this assumption in Section \ref{SS:RadHDiscussion}) and model the dust distribution of small grains by assuming a constant dust-to-gas mass ratio $f_\mathrm{dg}$ in the entire domain, thereby computing the dust density as $\rho_\mathrm{d} = f_\mathrm{dg}\, \rho$. We compute the absorption coefficients employing the same tabulated values as in \cite{Krieger2020,Krieger2022}, which are obtained for spherical grains composed of 62.5\% silicate and 37.5\% graphite with a density of $2.5$ g cm$^{-3}$ and a size distribution $\mathrm{d}n\sim a^{-3.5} \mathrm{d}a$ with radii in the range $a\in[5,250]$ nm.  We use $132$ $\kappa_\nu$ values logarithmically sampled in the frequency range  $[\nu_\mathrm{min},\nu_\mathrm{max}]=[1.5\times 10^{11},6\times 10^{15}]$ Hz, shown in Fig. \ref{fig:opacities}.

The radiation-matter interaction terms $(G^0,\mathbf{G})$
are computed by boosting into the laboratory frame
(see Eq. \eqref{Eq:SourceTerms}) their comoving values, given by
\begin{equation}\label{Eq:Gcomov}
  \begin{split}
  	\tilde{G}^0&=\kappa_P\,\rho_\mathrm{d}\left(\tilde{E}_r- a_R T^4\right) \\
  	\tilde{\mathbf{G}}&=\kappa_R\,\rho_\mathrm{d}\,\tilde{\mathbf{F}_r}\,,
  \end{split}
\end{equation}
where tildes indicate comoving quantities. In this expression
$a_r=4 \sigma_\mathrm{SB}/c$ is the radiation constant, $\sigma_\mathrm{SB}$ is the Stefan-Boltzmann constant, and $\kappa_P$ and $\kappa_R$ are, respectively, the Planck- and Rosseland-averaged absorption opacities. We thereby neglect scattering as an approximation. The opacity coefficients are computed as functions of the gas temperature $T$, as
\begin{equation}
    \kappa_P = \frac{
    \int_{\nu_\mathrm{min}}^{\nu_\mathrm{max}} \kappa_\nu
    \, B_\nu(T)\, \mathrm{d}\nu
    }{
    \int_{\nu_\mathrm{min}}^{\nu_\mathrm{max}}
    \, B_\nu(T)\, \mathrm{d}\nu
    }\,,\,\,\,
    \kappa_R^{-1}= \frac{
    \int_{\nu_\mathrm{min}}^{\nu_\mathrm{max}} \kappa_\nu^{-1}
    \, \frac{\partial B_\nu(T)}{\partial T}\, \mathrm{d}\nu
    }{
    \int_{\nu_\mathrm{min}}^{\nu_\mathrm{max}}
    \,  \frac{\partial B_\nu(T)}{\partial T}\, \mathrm{d}\nu
    }\,,
\end{equation}
where we use the same $\kappa_\nu$ values as for stellar irradiation.
The resulting values as a function of $T$ are shown in Fig. \ref{fig:opacities}. 
The gas temperature is computed according to the ideal
gas law
\begin{equation}\label{Eq:TempIdealGas}
T = \frac{\mu u}{k_\mathrm{B}}\frac{p}{\rho}\,,
\end{equation}
where $\mu$ and $u$ are respectively the gas mean molecular
weight and the atomic mass unit.
We assume immediate temperature
equilibrium between dust and gas,
and therefore this expression holds as
well for the temperature of dust grains.
We also neglect dust sublimation
and gas opacities,
since the temperature in our simulations
remains everywhere well below the $1000-1600$ K required for silicate
sublimation at our considered densities
\citep{IsellaNatta2005} and the $\sim 1500$ K required for the sublimation of carbon grains \citep{Duschl1996}. 

We integrate Eq. \eqref{Eq:RadHD} by evolving matter and radiation fields separately in an operator-split way, using substepping for the latter.  We employ implicit-explicit (IMEX) schemes for the explicit integration of radiation fluxes and the implicit integration of interaction terms, the last of which is carried out by means of the implicit method described in Appendix \ref{A:ImplicitMethod}.

\section{Construction of hydrostatic disk models}\label{S:DiskModels}

We construct our models of passive irradiated protoplanetary disks
by means of an iterative procedure based on \cite{Flock2013},
which yields density and temperature distributions under the assumption of hydrostatic equilibrium. Starting with a guess for both distributions, we update the temperature via two-moment radiative transfer (Section \ref{SS:HydroEqRadHD}), and recompute the density by solving hydrostatic equilibrium equations (Section \ref{SS:HydroEq}). These steps are then
repeated a number of times, using the previously obtained density configurations to compute new equilibrium temperatures. A convergence criterion is not applied to the resulting distributions since this process does not always lead to convergence, as we show in Section \ref{SS:ResultingModels}, where we analyze the obtained solutions with different parameters. In Section \ref{SS:RelaxationTime} we make an analysis of the disk's thermal relaxation timescale which is useful to later interpret the results of Section \ref{S:ResultsRadHD}.

\subsection{Temperature update}\label{SS:HydroEqRadHD}

We update the temperature distribution by modifying the Rad-HD module in order to describe radiative transfer in a stationary gas distribution, thus neglecting changes in the gas density and momentum. This results in the following system of equations:
 \begin{equation}\label{Eq:RadEqs}
 \begin{split}
\frac{\partial (\rho \epsilon)}{\partial t} &= c\,G^0  
  -\nabla\cdot\mathbf{F}_\mathrm{Irr} \\
\frac{1}{\hat{c}}\frac{\partial E_r}{\partial t}+\nabla\cdot \mathbf{F}_r &= -G^0 \\
\frac{1}{\hat{c}}\frac{\partial \mathbf{F}_r}{\partial t}+\nabla\cdot \mathbb{P}_r &= -\mathbf{G}\,,
\end{split}
\end{equation}
together with the equations of constant mass and momentum densities.
%These equations are used to compute the evolution of the gas internal energy density, $\rho \epsilon$, taking into account the balance between the heating due to stellar irradiation at the star temperature and the cooling and heating produced through radiative transfer at the gas temperature.
Equation \eqref{Eq:RadEqs} is solved by only applying the radiation substeps in the Strang-split method implemented in the code, and ignoring the hydro step \citep[see][]{MelonFuksman2021}. The implicit integration scheme in Appendix \ref{A:ImplicitMethod} is also modified in such a way that the momentum density is not updated. 

We assume assume a vertically isothermal disk in local thermal equilibrium (LTE) as initial condition for
the first iteration and integrate Eq. \eqref{Eq:RadEqs} 
for a total time $t_\mathrm{iter}$ per iteration.
After updating the density field for a new run,
the fields $\rho \epsilon$, $E_r$, and $\mathbf{F}_r$
are initialized to their final values at the previous
iteration.

We assume for the computation of the irradiation flux the same stellar parameters as e.g. \cite{Watanabe2008,Ueda2021}, corresponding to a T Tauri star with $T_s=4000$ K, $Ms=M_\odot$, and $Rs=2.086 R_\odot$, in such a way that the total luminosity is $L_\odot$. We take $\Gamma=1.41$, which is typical for solar composition \citep[][]{DeCampli1978}. We define the computational domain as $(r,\theta)\in[0.4,100]\,\mathrm{au}\times[\pi/2-0.3,\pi/2+0.3]$, and assume $r_0=10\,R_s$ for the computation of $\tau_0$ (Appendix \ref{A:OptDepth}). Our discretization uses $N_r\times N_\theta=512\times 200$ cells logarithmically and uniformly distributed in the $r$-- and $\theta$--directions, respectively, in such a way that the cell aspect ratio is always $\Delta r/r\Delta \theta\approx 3.6$. We employ zero-gradient boundary conditions for the radiation fields at the radial boundaries, and prevent radiation inflow by reflecting the radiation flux at the ghost cells if it is directed towards the domain. The same is done with the flux at the $\theta$-boundaries, while $E_r$ is set to a small value given by $E_r=a_R\,(10\,\mathrm{K})^4$ at the $\theta$-ghost cells to guarantee that radiation escapes the domain. We use in every case second-order piecewise linear reconstruction with the modified Van Leer-like limiter by \cite{Mignone2014reconstruction}, and integrate Eq. \eqref{Eq:RadEqs} by means of the IMEX1 scheme in \cite{MelonFuksman2019} which extends the second-order total variation diminishing (TVD) Runge-Kutta method by \cite{GottliebShu1996}. We use the HLLC Riemann solver by \cite{MelonFuksman2019} to compute the intercell radiation fluxes.

\subsection{Hydrostatic equilibrium}\label{SS:HydroEq}

Given each temperature distribution, we obtain radial and
vertical hydrostatic equilibrium conditions
by setting all time derivatives to zero in Eq.
\eqref{Eq:RadHD}, as well as $v_r$, $v_\theta$.
Additionally, we neglect radiative momentum exchange by setting $\mathbf{G}$ to zero in the momentum equation
(second in Eq. \eqref{Eq:RadHD}), which
is justified given that the components of $\mathbf{G}$ in all of our simulations are everywhere smaller than those of
$\nabla \Phi$, $\nabla p$, and $\nabla\cdot(\rho\mathbf{v}\mathbf{v})$ by several orders of magnitude.
Under these assumptions, the $(r,\theta)$-components of the
momentum equation become
\begin{equation}\label{Eq:Hydrostatic}
    \begin{split}
        \frac{\partial p}{\partial r} &=
        -\rho \frac{\partial \Phi}{\partial r}
        + \frac{\rho v_\phi^2}{r}  \\
        \frac{1}{r} \frac{\partial p}{\partial \theta}
        & = \frac{1}{\tan \theta}  \frac{\rho v_\phi^2}{r}\,.
    \end{split}
\end{equation}
These equations can be rewritten in terms of $\rho$, $v_\phi$ ,and $\overline{T}=k_B T/\mu u$
by means of Eq. \eqref{Eq:TempIdealGas}, yielding
\begin{equation}\label{Eq:Hydrostatic2}
    \begin{split}
         \frac{v_\phi^2}{r}  &=
         \frac{\partial \Phi}{\partial r}
         + \frac{\partial \overline{T}}{\partial r}
         + \overline{T} \frac{\partial \log \rho}{\partial r}
         \\ 
         \frac{\partial \log \rho}{\partial \theta} &=
        \frac{1}{\tan{\theta}}
       \frac{v_\phi^2}{\overline{T}}
        -\frac{1}{\overline{T}}\frac{\partial \overline{T}}{\partial \theta}\,.
    \end{split}
\end{equation}
The only unknowns in these equations are $\rho$ and $v_\phi$, which can be computed by defining $\rho$ for all $r$ at a given $\theta=\theta_0$. The second equation in Eq. \eqref{Eq:Hydrostatic2}
can then be integrated in the $\theta$-direction starting from $\theta_0$,
computing $v_\phi$ at every step via the first equation. We use for this integration a second-order Runge Kutta method (midpoint rule).
We assume the following functional form for $\rho(r,\theta)$ at $\theta_0=\pi/2$, valid for thin, vertically isothermal disks:
\begin{equation}\label{Eq:rhomidplane}
    \rho(r,\pi/2)=\frac{\Sigma_0(R)}{\sqrt{2\pi}H(R)}\,,
\end{equation}
where $\Sigma_0(R)$ is the initial vertically integrated density, which
is kept constant among iterations, while $H(R)=c_s(R)/\Omega_K$ is the  midplane pressure scale height, where $R=r \sin \theta$ is the cylindrical radius.
The latter is defined in terms of the midplane isothermal sound speed $c_s(R)=\sqrt{\overline{T}(r,\pi/2)}$
and the Keplerian angular velocity $\Omega_K=\sqrt{G\,M_s/r^3}$.
Even though the disk is not everywhere vertically isothermal
close to the midplane (later shown in Fig. \ref{fig:Ttheta}), where most of the mass is located, the surface density computed by vertically integrating Eq. \eqref{Eq:rhomidplane} stays close to $\Sigma_0(R)$
within $1$\% in all iterations.

We assume $\Sigma_0(R)\propto R^{-1}$ as in \cite{WuLitwick2021}, which is shallower than in the minimum-mass solar nebula model by \cite{Hayashi1981} considered in other works on this topic \citep[e.g.,][]{Watanabe2008} but consistent with observed disks around T Tauri stars. We take $\Sigma_0(R)=200\,\mathrm{g\,cm^{-2}}\,(R/\mathrm{au})^{-1}$, which is  similar to the distribution estimated e.g. in \cite{VanBoekel2017} and an order of magnitude below that in \cite{Watanabe2008,Ueda2021} at $1$ au. As in \cite{WuLitwick2021}, we define the initial temperature distribution in such a way that the initial aspect ratio is $H/R=0.0248 (R/\mathrm{au})^{0.275}$ taking $\mu=2.3$.

For the integration of Eq. \eqref{Eq:Hydrostatic2}, we impose a density floor of $\rho_\mathrm{min}=10^{-27}$ g cm$^{-3}$ to avoid
errors in small-number operations when integrating Eq. \eqref{Eq:Hydrostatic2} over several orders of magnitude in density. Since this is only applied in optically thin regions of space at the disk atmosphere, it does not affect the computed temperatures, as these are independent from the density for $\tau\ll 1$.

\subsection{Resulting models}\label{SS:ResultingModels}

We obtained a series of hydrostatic configurations characterized by the iteration time, the dust-to-gas mass ratio, and the reduced speed of light. As summarized in Table \ref{tt:hydrost}, simulations are labeled following the nomenclature \sftw{dgN$_\mathrm{dg}$tN$_{t}$}, where $N_\mathrm{dg}$ and $N_{t}$ are defined as $f_\mathrm{dg}=10^{-N_\mathrm{dg}}$ and $t_\mathrm{iter}=N_\mathrm{t}$ yr, respectively. We considered $f_\mathrm{dg} = 10^{-4}$, $10^{-3}$, and $10^{-2}$. A vertically integrated mass ratio of $10^{-2}$ spanning all grain sizes is typical for protoplanetary disks, while the small dust grains can typically contribute between $1$ and $10\%$ of the total dust mass \citep[see, e.g.,][]{Birnstiel2012}. We then regard $f_\mathrm{dg}=10^{-2}$ as a case of extremely high dust content, which we consider in order to study the disk stability in a regime that favors self-shadowing. We use $\hat{c}/c=10^{-4}$ in most cases, and otherwise indicate this value with an additional \_cN$_c$ label ($\hat{c}/c=10^{-N_c}$).

\begin{table}[t!]
\centering
\caption{Parameters of shown hydrostatic simulations: dust-to-gas mass ratio $f_\mathrm{dg}$, time $t_\mathrm{iter}$ per iteration for radiative transfer, and $\hat{c}/c$ ratio between the reduced and real values of the speed of light.}
\label{tt:hydrost}
\begin{tabular}{*4l}
\hline\hline
Label              & $f_\mathrm{dg}$ & $t_\mathrm{iter}$ (yr) & $\hat{c}/c$ \\ \hline
\sftw{dg2t100}     & $10^{-2}$       & 100                    & $10^{-4}$   \\
\sftw{dg2t10}      & $10^{-2}$       & 10                     & $10^{-4}$   \\
\sftw{dg2t1}       & $10^{-2}$       & 1                      & $10^{-4}$   \\
\sftw{dg2t1\_c3}   & $10^{-2}$       & 1                      & $10^{-3}$   \\
\sftw{dg2t1\_c2}   & $10^{-2}$       & 1                      & $10^{-2}$   \\
\sftw{dg2t0.1}     & $10^{-2}$       & 0.1                    & $10^{-4}$   \\
\sftw{dg3t100}     & $10^{-3}$       & 100                    & $10^{-4}$   \\
\sftw{dg3t10}      & $10^{-3}$       & 10                     & $10^{-4}$   \\
\sftw{dg3t1}       & $10^{-3}$       & 1                      & $10^{-4}$   \\
\sftw{dg3t0.1}     & $10^{-3}$       & 0.1                    & $10^{-4}$   \\
\sftw{dg4t100}     & $10^{-4}$       & 100                    & $10^{-4}$   \\
\sftw{dg4t10}      & $10^{-4}$       & 10                     & $10^{-4}$   \\
\sftw{dg4t1}       & $10^{-4}$       & 1                      & $10^{-4}$   \\
\sftw{dg4t0.1}     & $10^{-4}$       & 0.1                    & $10^{-4}$   \\ \hline
\end{tabular}
\end{table}

\begin{figure*}[t!]
\centering
\includegraphics[width=\linewidth]{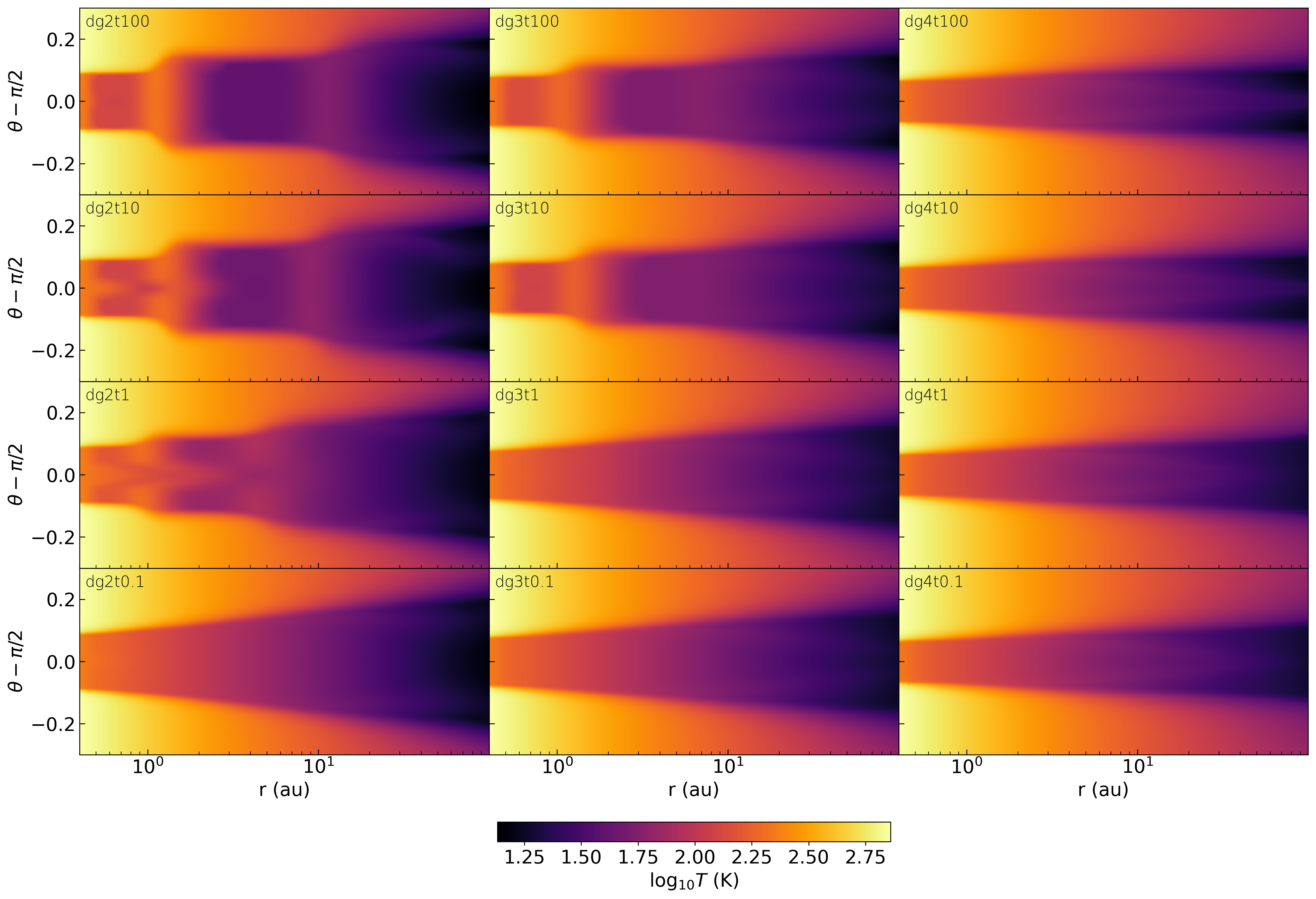}
\caption{2D temperature profiles in log scale for all hydrostatic runs with $\hat{c}=10^{-4}$. Distributions are shown after $35$ iterations for $t_\mathrm{iter}=100$, $10$, and $1$ yr (top three rows) and $400$ iterations for $t_\mathrm{iter}=0.1$ (bottom row). Videos made with snapshots of these runs at different iterations for $t_\mathrm{iter}=1$ and $100$ yr can be found in \href{https://youtu.be/RT8IFe8W13g}{https://youtu.be/RT8IFe8W13g}.}
\label{fig:T2D}
\end{figure*}

\begin{figure*}[t!]
\centering
\includegraphics[width=\linewidth]{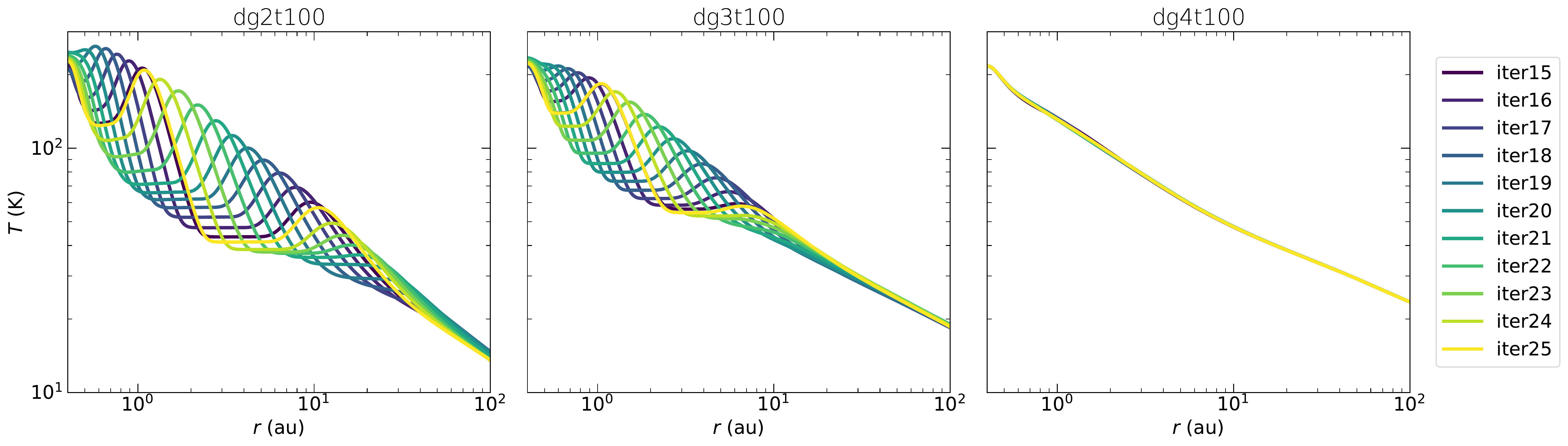}
\caption{Midplane temperature profiles in the hydrostatic runs between iterations $15$ and $25$ for each $f_\mathrm{dg}$.}
\label{fig:Titer}
\end{figure*}

\begin{figure*}[t!]
\centering
\includegraphics[width=\linewidth]{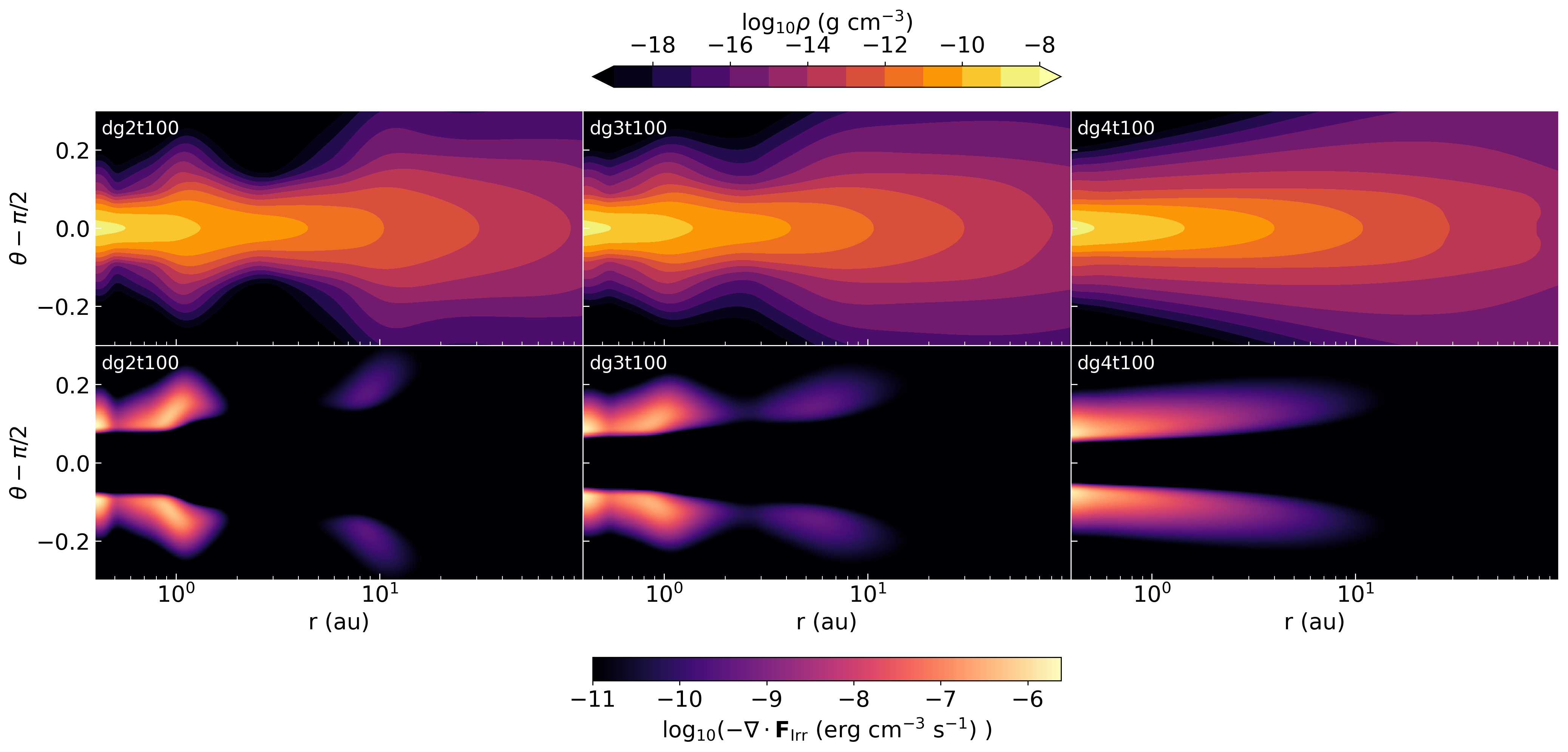}
\caption{Density (top) and irradiation power (bottom) hydrostatic distributions after 35 iterations for each $f_\mathrm{dg}$ in the $t_\mathrm{iter}=100$ yr runs.}
\label{fig:rhofir2D}
\end{figure*}

Figure \ref{fig:T2D} shows the resulting temperature distributions for all employed $f_\mathrm{dg}$ and $t_\mathrm{iter}$. Results are shown after $35$ iterations for $t_\mathrm{iter}\geq 1$ yr and after $400$ for $t_\mathrm{iter}= 0.1$ yr. In the latter case, we do this to give the system enough time to depart from the initial condition and reach a state that repeats itself after a certain number of iterations. We focus for now on the longest $t_\mathrm{iter}$ runs and consider the rest of them in Section \eqref{SS:RelaxationTime}. In such cases, the solutions for $f_\mathrm{dg}=10^{-2}$ and $10^{-3}$ do not converge, and instead exhibit peaks or "bumps" in the midplane temperature which are more prominent for higher dust content. For $f_\mathrm{dg}=10^{-4}$, the solutions converge to a constant distribution after $12$ iterations. As shown in Fig. \ref{fig:Titer}, temperature bumps in the $t_\mathrm{iter}=100$ yr solutions are continuously formed and shifted towards the star among iterations, reaching the same configuration every $10$ iterations. This saturated behavior is reached after $10$ iterations and maintained thereafter without further increasing the maximum amplitude of the temperature peaks. As mentioned in the introduction, the radial shifting is caused by the heating and expansion of the bump's inner side, which is exposed to the star, and the cooling and contraction of the shadowed regions. This asymmetric heating can be seen in Fig. \ref{fig:rhofir2D}, where density contours are shown together with power irradiation maps for $t_\mathrm{iter}=100$ yr. For $f_\mathrm{dg}=10^{-4}$, the entire disk surface is illuminated by the star, and there are no hints of such shadowing after convergence. 

\begin{figure*}[t!]
\centering
\includegraphics[width=\linewidth]{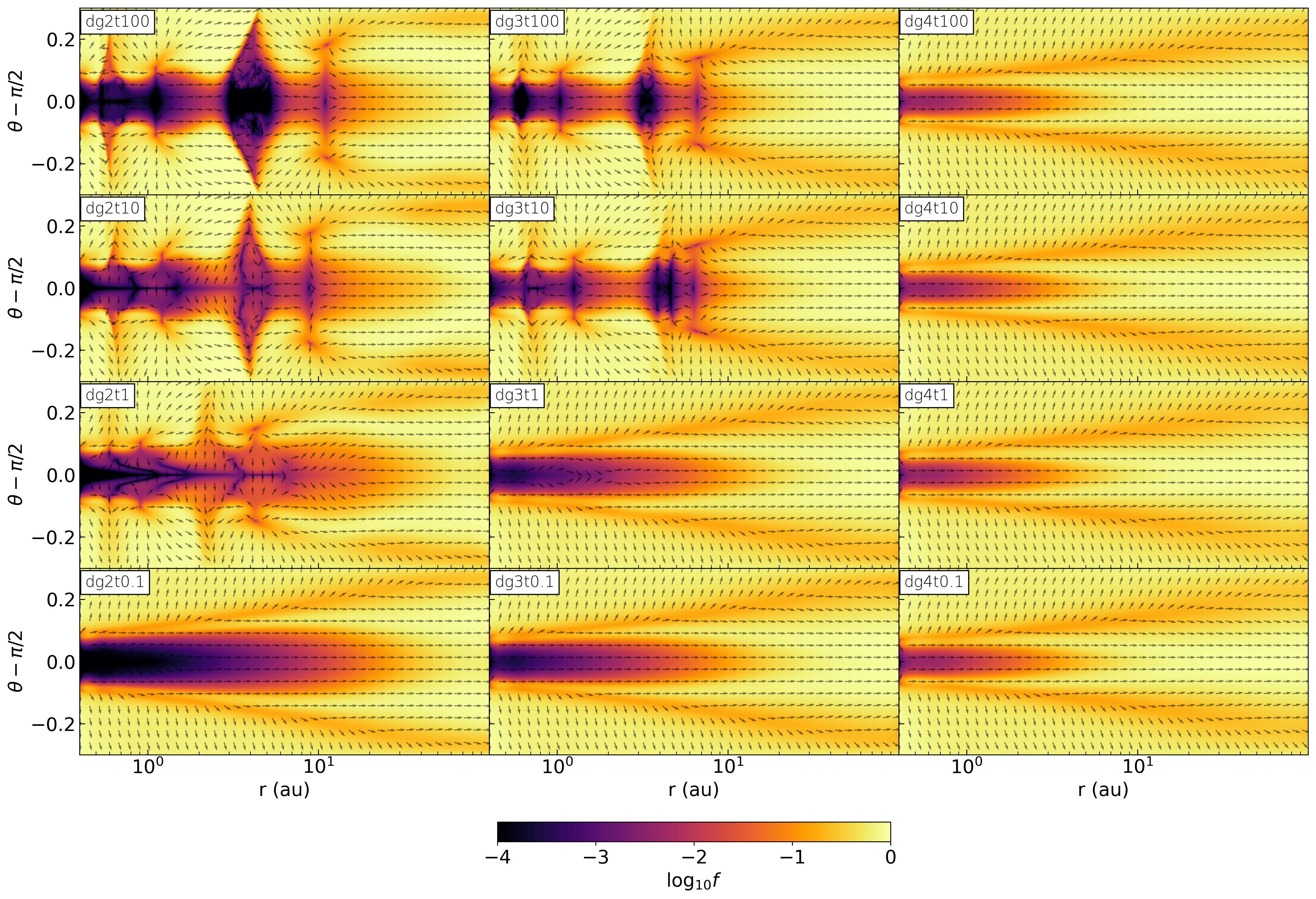}
\caption{Same as Fig. \ref{fig:T2D} showing color maps of the reduced flux $f=||\mathbf{F}_r||/E_r$ in log scale. Arrows indicate the direction of $\mathbf{F}_r$.}
\label{fig:f2D}
\end{figure*}

As in \cite{WuLitwick2021} and \cite{Okuzumi2022}, we do not observe a formation of scale height perturbations at a few tenths of au. The reason for this is that the disk is optically thin in such regions, and any local temperature perturbation is immediately damped by radial radiation transport. This can be noted by taking a look at the distribution of the reduced flux $f=||\mathbf{F}_r||/E_r$, which is much smaller than $1$ in the diffusion regime, particularly in the shadowed regions, and tends to $1$ in the free-streaming regime. We show this quantity and the flux direction in Fig. \eqref{fig:f2D}. For small radii ($\lesssim10$ au), the radiation flux transitions between the diffusion regime in the disk interior and free streaming in the optically thin atmosphere. In the transition layer between both regimes, the flux emanates predominantly from the overheated regions, and escapes the domain through the $\theta$-boundary. At larger radii, radiation freely streams in the radial direction at the midplane. For $f_\mathrm{dg}=10^{-2}$ and $10^{-3}$, temperature bumps only start forming in the diffusion-dominated optically thick region of the disk. The extent of this region decreases for smaller $f_\mathrm{dg}$, ending at approximately $30$, $10$, and $2$ au for $f_\mathrm{dg}=10^{-2}$, $10^{-3}$, and $10^{-4}$ respectively. In the latter case, the disk is not optically thick enough for self-shadowed temperature peaks to form.
% In the latter case, the extent of the optically thick region is not large enough for self-shadowed temperature peaks to form.
Moreover, the radial radiation flux at the midplane increases for smaller $f_\mathrm{dg}$, with a minimum $f$ scaling roughly as $3\times 10^{-7}/f_\mathrm{dg}$.

\begin{figure*}[t!]
\centering
\includegraphics[width=\linewidth]{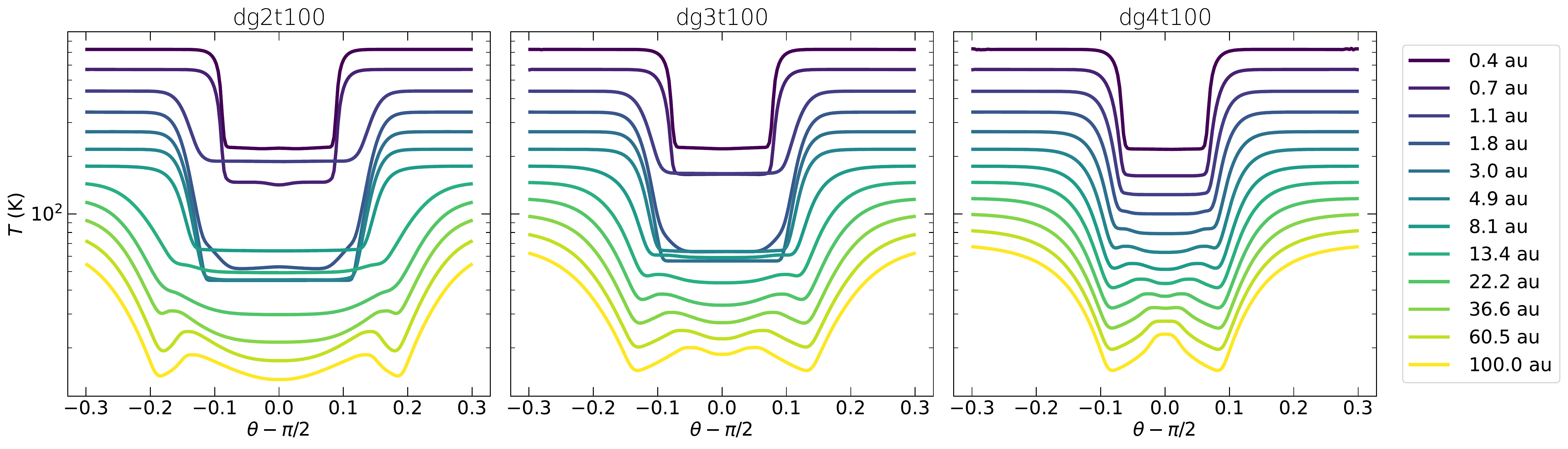}
\caption{Hydrostatic $T(\theta)$ profiles at different radii for each $f_\mathrm{dg}$, computed after $35$ iterations with $t_\mathrm{iter}=100$ yr.}
\label{fig:Ttheta}
\end{figure*}

\begin{figure*}[t!]
\centering
\includegraphics[width=\linewidth]{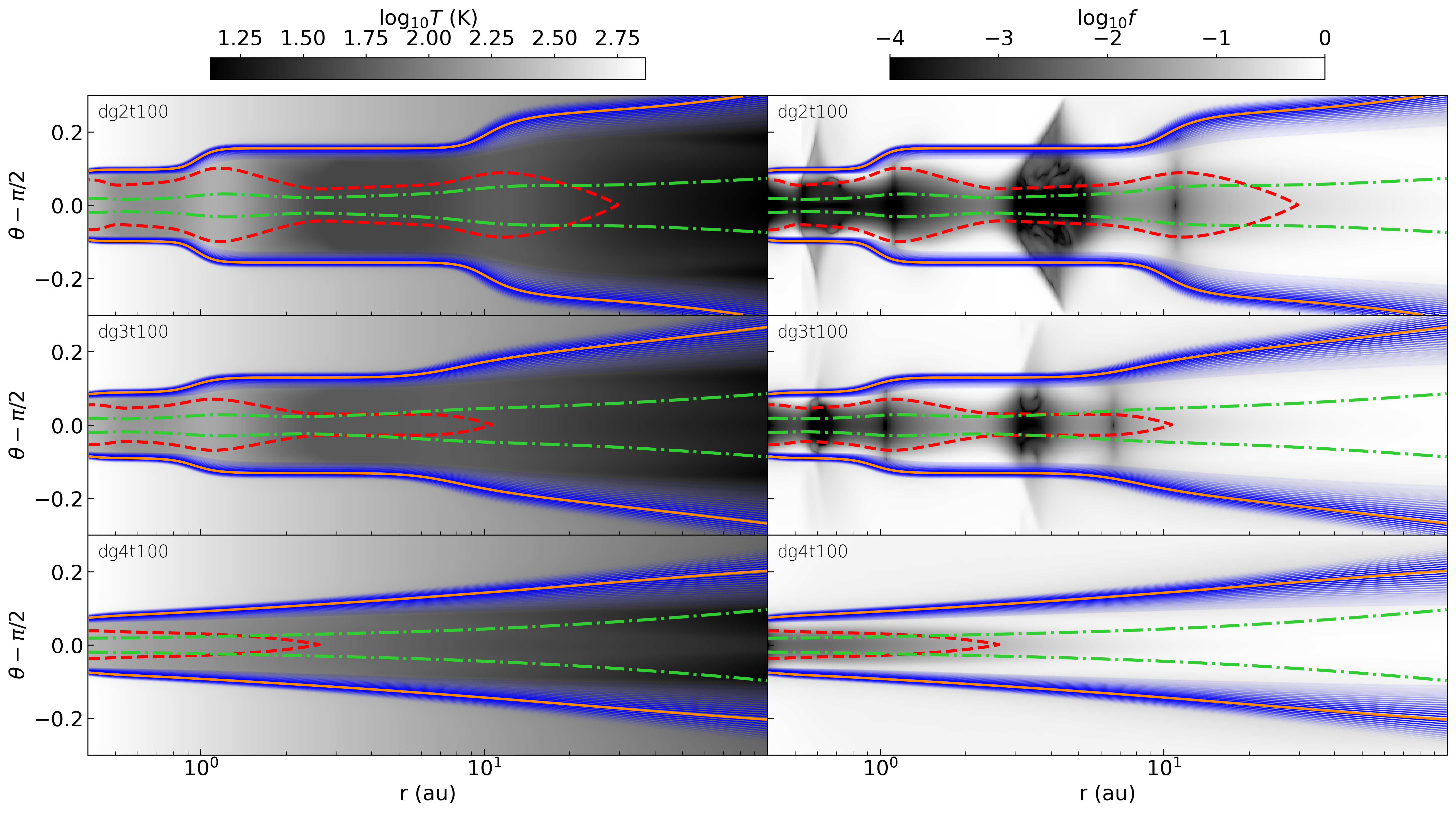}
\caption{Planck-averaged $\tau=1$ surfaces for vertical radiative transport (dashed red lines), frequency-dependent $\tau=1$ surfaces integrated along $r$ for stellar irradiation (thin blue lines), Planck-averaged $\tau=1$ surfaces for stellar irradiation (thick orange lines), and pressure scale height $z=H$ surfaces (green dash-dotted lines) in the hydrostatic runs with $t_\mathrm{iter}=100$ yr. The temperature and reduced radiation flux distributions are represented in grayscale for comparison. The alpha factor used for the transparency of the $\tau=1$ lines is proportional to the contribution of each frequency to the total irradiation power.}
\label{fig:2Dtausurfaces}
\end{figure*}

The boundary between the optically thick and optically thin regions is approximately delimited by the surfaces of unity vertical optical depth ($\tau=1$) computed integrating $\kappa_P(T) \rho_\mathrm{d} \mathrm{d}z$ from the upper and lower boundaries. The location of these surfaces is shown in Fig. \ref{fig:2Dtausurfaces} for all simulations with $t_\mathrm{iter}=100$ yr, together with the $T$ and $f$ distributions. The scale height $z=H(R)$ and the irradiation $\tau=1$ surfaces are shown on the same figure, where the latter is computed for each frequency using Eq. \eqref{Eq:OpticalDepth}. It can be seen that the ratio $z_s/H$ between the approximate height of the irradiation layer and $H$ varies at the temperature bumps, which is the condition for instability derived in \cite{WuLitwick2021}. This ratio increases with $f_\mathrm{dg}$, which is argued in the same work to favor the instability.

We note in these figures that the obtained $f$ values decrease near the irradiation surfaces due to the local increase of $E_r$. Above these surfaces, radiation is transported out of the domain in the vertical direction. Below, part of the energy heats up the optically thick disk regions, and the rest is transported radially through the optically thin atmosphere. In the optically thin regions at large radii, this radial flux converges with the flux towards the midplane caused by irradiation heating. This convergence explains why the $T(\theta)$ profiles depart in such regions from the approximately vertically isothermal distribution verified in the optically thick regions, as shown in Fig. \ref{fig:Ttheta}. This phenomenon is a direct consequence of solving radiative transport using a fluid approach, since convergent fluxes do not interact in a real scenario. More refined radiative transport schemes, such as methods relying on an angular discretization of the radiation specific intensity \citep[see, e.g.,][]{Davis2012,Jiang2021}, should be able to yield more accurate radiation fluxes at large radii. However, this effect has a small impact on the radiation flux and the disk temperature, and most importantly, it does not alter the conclusions of this paper.

\subsection{Thermal relaxation time}\label{SS:RelaxationTime}

\begin{figure}[t!]
\centering
\includegraphics[width=\linewidth]{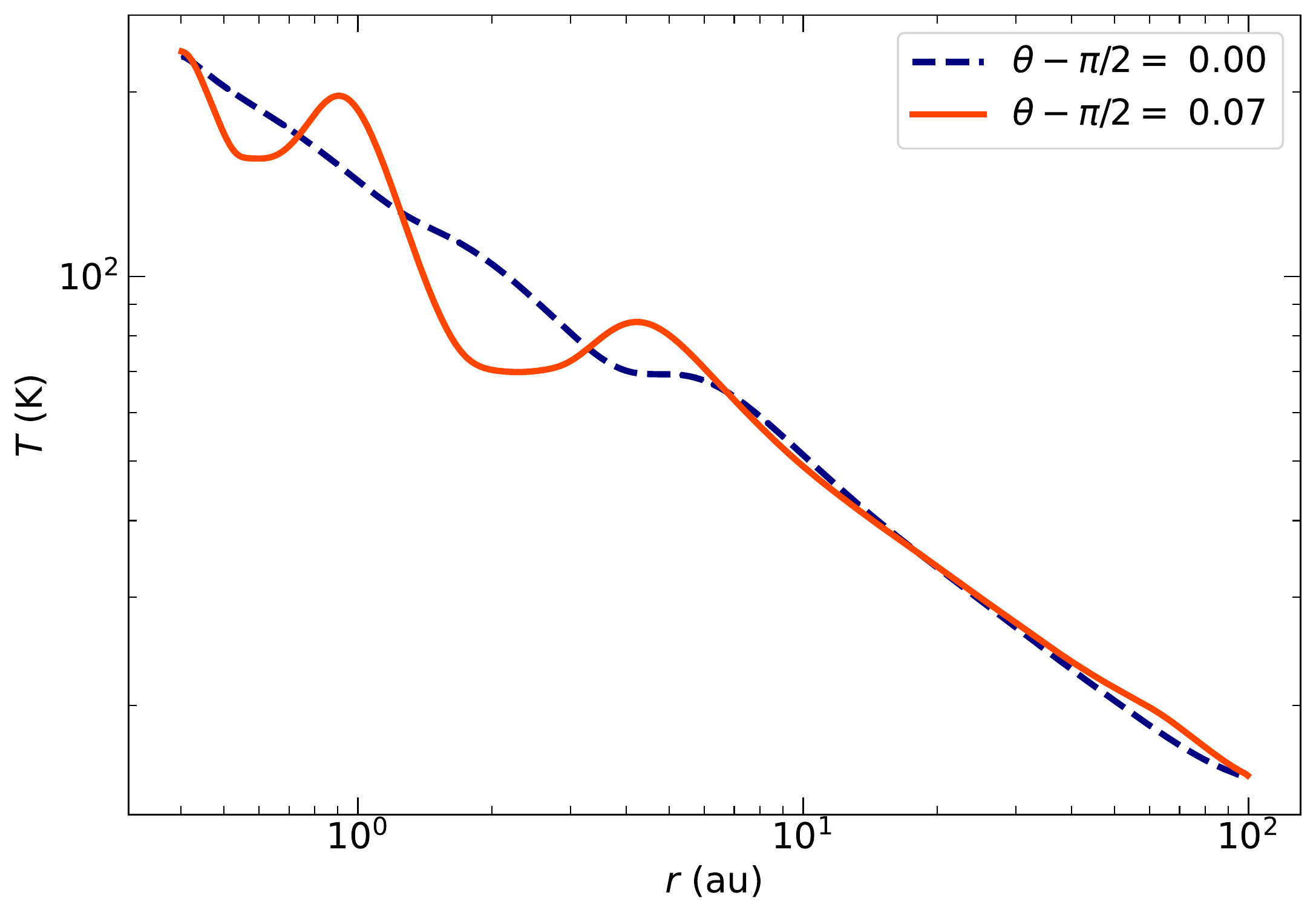}
\caption{Radial temperature profiles at $\theta-\pi/2=0$ and $0.07$ in run \sftw{dg2t1\_c2} after $35$ iterations.}
\label{fig:T1D_h}
\end{figure}

\begin{figure}[t!]
\centering
\includegraphics[width=\linewidth]{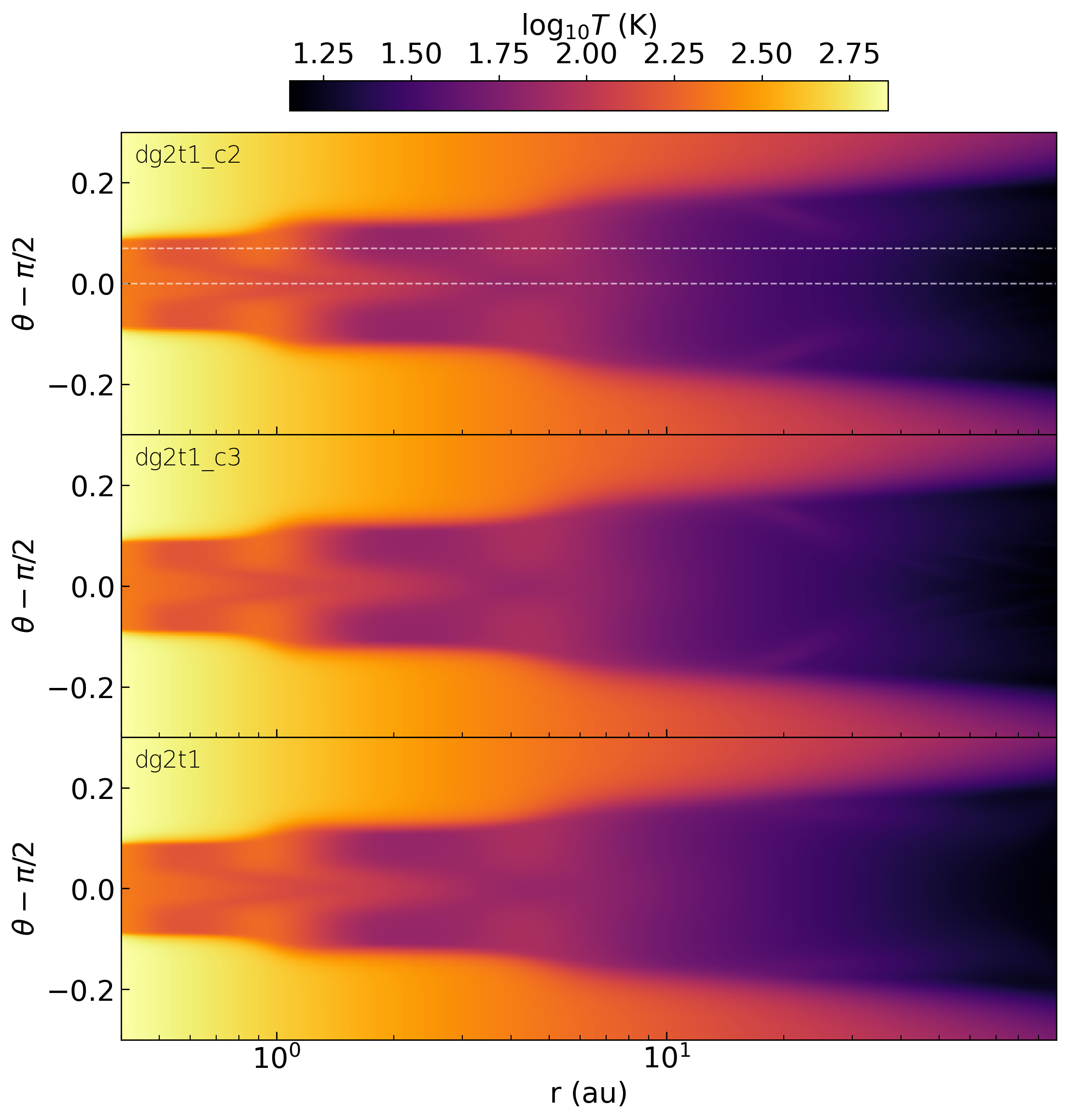}
\caption{Temperature distributions after 35 iterations for $f_\mathrm{dg}=10^{-2}$, $t_\mathrm{iter}=1$ yr, and $\hat{c}=10^{-2}$ (top), $\hat{c}=10^{-3}$ (middle), and $\hat{c}=10^{-4}$ (bottom). The horizontal white dashed lines indicate the $\theta$ values where the temperature profiles in Fig. \ref{fig:T1D_h} were taken.}
\label{fig:T2D_c}
\end{figure}

Figure \ref{fig:T2D} shows that the shadowing-induced temperature perturbations are not vertically uniform for sufficiently short $t_\mathrm{iter}$. The temperature distributions shown for runs \sftw{dg2t10} and \sftw{dg2t1} evidence that, at the end of each iteration, not enough time has passed to adjust the midplane temperature to a new density configuration. This occurs because most of the stellar irradiation is absorbed far from the midplane at the irradiation layers. After each density update, the regions close to those surfaces are updated first, and the temperature is adjusted to the new configuration via diffusion from the outer layers to the midplane. As a consequence, if $t_\mathrm{iter}$ is short enough, there is a middle layer close to the midplane where the temperature has not yet adjusted. Naturally, the thickness of that middle layer grows when $t_\mathrm{iter}$ is reduced, since that means that there is less time for temperature perturbations to diffuse into the disk. We can also see that the temperature peaks at the midplane are located at larger radii with respect to those at the surface, as they trace the temperature maxima at the previous iterations. In these cases, the temperature perturbations at the midplane are smaller than at the surface, as can be seen in Fig. \ref{fig:T1D_h}, where we show the temperature in run \sftw{dg2t1} as a function of radius at two different altitudes, $\theta-\pi/2=0$ and $0.07$, indicated on Fig. \ref{fig:T2D_c}. This occurs because thermal perturbations do not reach the midplane via diffusion before overheated regions become shadowed regions and vice versa, which averages out the cooling and heating produced at the midplane. This effect, combined with the growth of the middle layer as $t_\mathrm{iter}$ is reduced, causes temperature perturbations to almost entirely disappear for short enough $t_\mathrm{iter}$, as shown in Fig. \ref{fig:T2D} for \sftw{dg2t0.1} and \sftw{dg3t1}. For $t_\mathrm{iter}= 0.1$ yr, the maximum relative temperature changes between iterations are below $10^{-4}$ in all cases after $400$ iterations. It is likely that the reason why no outer perturbed layers remain for short enough $t_\mathrm{iter}$ is that these have shrunk enough to be entirely confined in optically thin regions, where temperature perturbations are damped out by radiative transport.

An important issue that must be addressed is if this time-dependent diffusion is affected by the employment of a reduced speed of light. Fig. \ref{fig:T2D_c} shows that this is not the case for our chosen $\hat{c}$ values. We show in this figure the temperature distributions for \sftw{dg2t1}, \sftw{dg2t1\_c3}, and \sftw{dg2t1\_c2} after $35$ iterations. We only observe slight differences in the optically thin region in the features caused by convergent radiation fluxes (Section \ref{SS:ResultingModels}) for $\hat{c}/c=10^{-4}$. In the optically thick region, the vertical extent and temperature distribution of the middle layer that is not efficiently heated through vertical diffusion are the same in all three cases. The reason for this is that the value of $\hat{c}$ controls the adjustment time of the radiation fields to each new density distribution, but disappears from the equations once these reach a quasistatic configuration, as can be seen by neglecting the terms $\frac{1}{\hat{c}}\partial_t E_r$ and $\frac{1}{\hat{c}}\partial_t\mathbf{F}_r$ in Eq. \eqref{Eq:RadEqs}. After a fast readjustment of the radiation fields, the cooling time of the disk is actually regulated by the $c\,G^{0}$ term on the right-hand side of Eq. \eqref{Eq:RadEqs}, which depends on the real value of $c$. We can therefore rely on the results of this section to estimate the characteristic cooling time $t_\mathrm{cool}$ in which surface temperature perturbations reach the midplane via diffusion. From Fig. \ref{fig:T2D}, we obtain $t_\mathrm{cool}\sim100$ yr for $f_\mathrm{dg}=10^{-2}$ and $\sim10$ yr for $f_\mathrm{dg}=10^{-3}$.

The just described effect is not seen in other works on this topic because of the assumption of instant information mentioned in the introduction. Even studies that replace a detailed modeling of time-dependent vertical diffusion with an increase of the cooling timescale, such as that in \cite{Okuzumi2022}, do not account for the delay between the surface and midplane temperature distributions. On the other hand, iterative methods relying on Monte Carlo simulations cannot exhibit features due to vertical diffusion in finite time since they correspond to the $t_\mathrm{iter}\rightarrow \infty$ limit, and can only be compared to our $t_\mathrm{iter}=100$ yr runs. In other words, our longest-$t_\mathrm{iter}$ runs mimic the result of assuming instantaneous thermal relaxation.

The results of this section suggest that a self-shadowing instability such as that described in literature can only be possible if density and temperature perturbations at the disk surface propagate towards the midplane faster than they change at the surface via either cooling or dynamical processes. Identifying which of these competing effects prevails requires a time-dependent analysis, such as the one we conduct in the next section.
%by computing the disk evolution in Rad-HD simulations.

\section{Time evolution}\label{S:ResultsRadHD}

\subsection{Rad-HD simulations}\label{SS:RadHDSims}

\begin{table}[t!]
\centering
\caption{Parameters of shown Rad-HD simulations: dust-to-gas mass ratio $f_\mathrm{dg}$ and $\hat{c}/c$ ratio between the reduced and real values of the speed of light.}
\label{tt:radhd}
\begin{tabular}{*3l}
\hline\hline
Label             & $f_\mathrm{dg}$ & $\hat{c}/c$ \\ \hline
\sftw{rhdg2\_c3}  & $10^{-2}$       &  $10^{-3}$  \\
\sftw{rhdg2\_c4}  & $10^{-2}$       &  $10^{-4}$  \\
\sftw{rhdg3\_c3}  & $10^{-3}$       &  $10^{-3}$  \\
\sftw{rhdg3\_c4}  & $10^{-3}$       &  $10^{-4}$  \\
\sftw{rhdg4\_c3}  & $10^{-4}$       &  $10^{-3}$  \\
\sftw{rhdg4\_c4}  & $10^{-4}$       &  $10^{-4}$  \\
\hline
\end{tabular}
\end{table}

\begin{figure*}[t!]
\centering
\includegraphics[width=\linewidth]{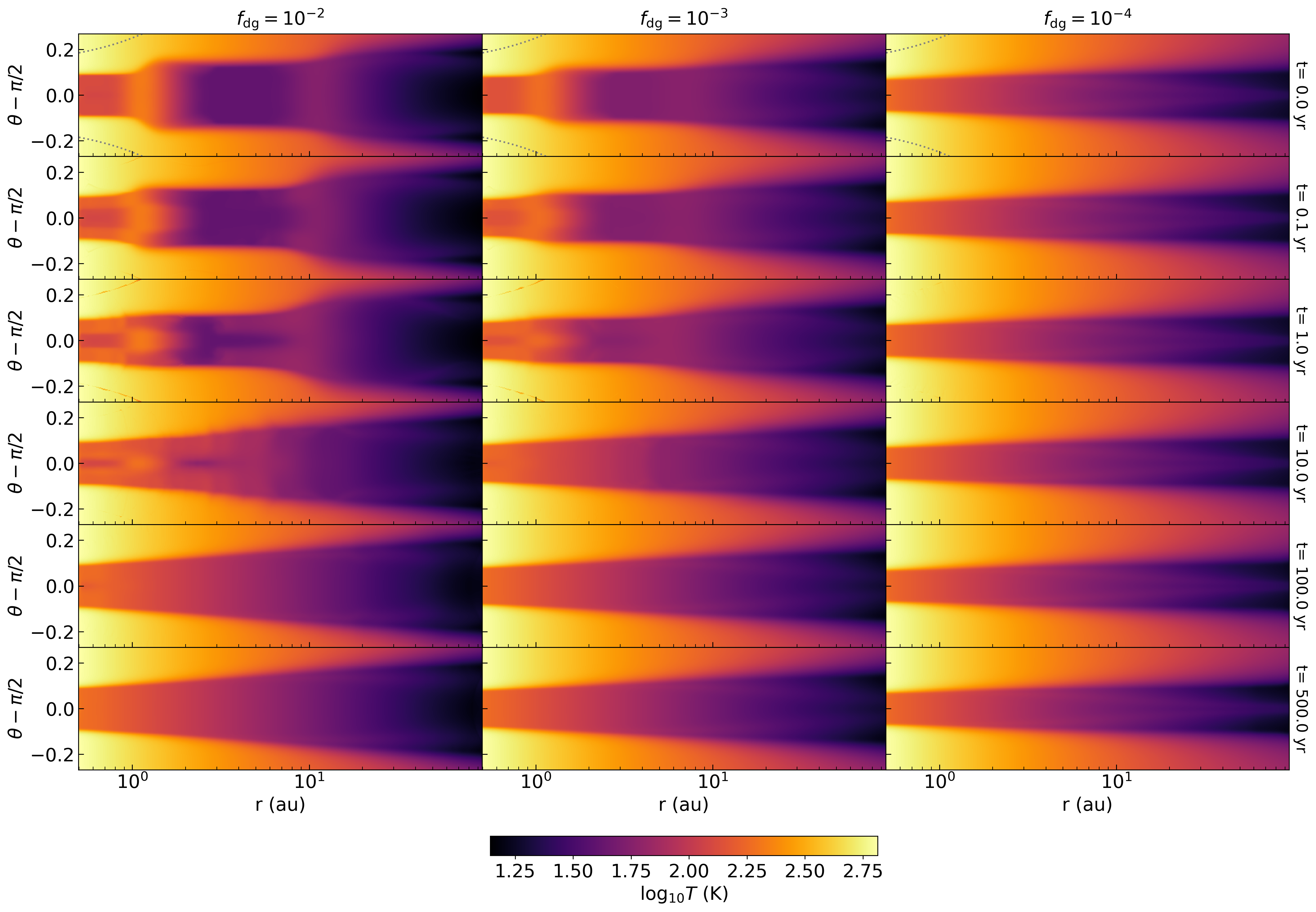}
\caption{Temperature distributions as a function of time for runs \sftw{rhdg2\_c3} (left), \sftw{rhdg3\_c3} (middle), and \sftw{rhdg4\_c3} (right). The boundaries of the damping zones are indicated with dotted lines in the $t=0$ snapshots. Videos made with snapshots of these runs can be found in \href{https://youtu.be/RT8IFe8W13g}{https://youtu.be/RT8IFe8W13g}.}
\label{fig:rh_T2D}
\end{figure*}

\begin{figure*}[t!]
\centering
\includegraphics[width=\linewidth]{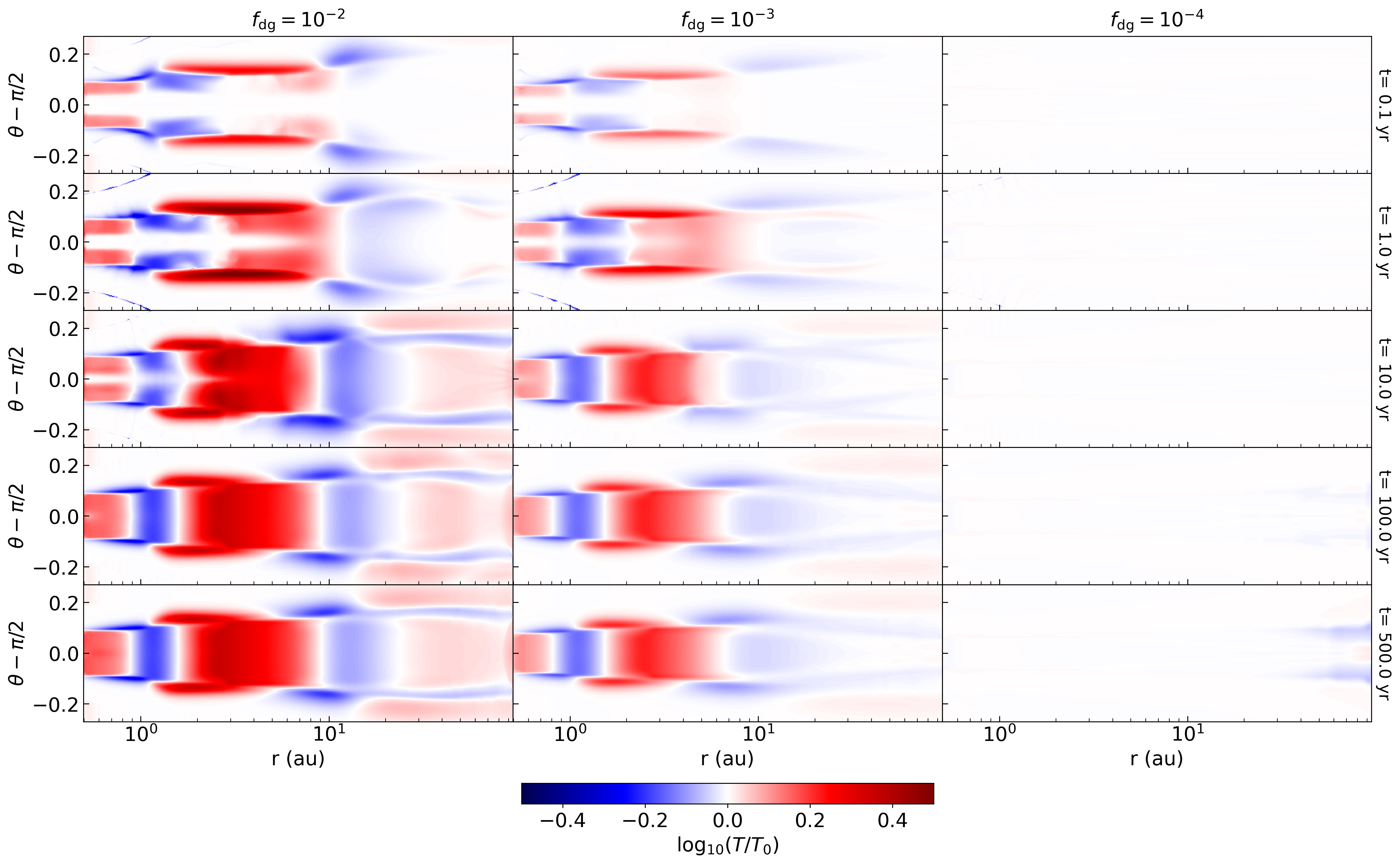}
\caption{Same as Fig. \ref{fig:rh_T2D}, this time showing a time-dependent comparison of the current temperature $T$ with the initial temperature $T_0$.}
\label{fig:rh_dT2D}
\end{figure*}

\begin{figure*}[t!]
\centering
\includegraphics[width=\linewidth]{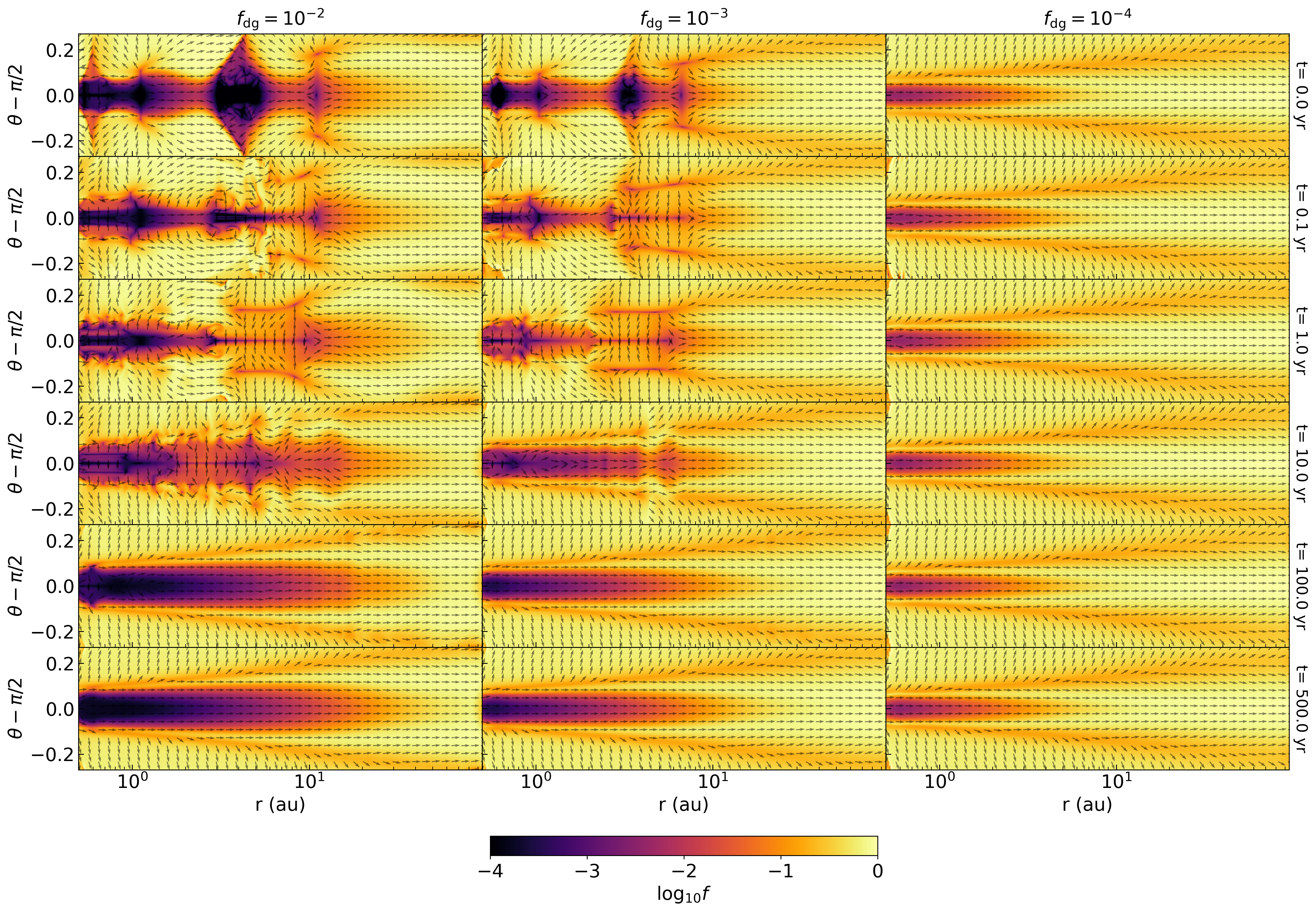}
\caption{Same as Fig. \ref{fig:rh_T2D}, showing the evolution of $\log_{10}f$ (color scale) and the direction of $\mathbf{F}_r$ (arrows). }
\label{fig:rh_f}
\end{figure*}

Until now we have verified that assuming hydrostatic equilibrium and instantaneous vertical diffusion naturally leads to the spontaneous formation of temperature bumps for high enough dust content. We now focus on whether such temperature perturbations survive and keep forming if gas advection is switched back on. With this purpose, we ran Rad-HD simulations starting from the longest-$t_\mathrm{iter}$ solutions obtained in the previous section. Given the suggestion by \cite{WuLitwick2021} that iterative procedures like the one used to construct such solutions might enhance the formation of temperature perturbations in comparison with time-dependent simulations, we can regard these initial conditions as the most favorable case for the development of a self-shadowing instability.

We summarize the list of our time-dependent Rad-HD runs in Table \ref{tt:radhd}. Simulations are characterized by the employed values of $f_\mathrm{dg}$ and $\hat{c}/c$ and labeled as \sftw{rhdgN$_\mathrm{dg}$\_cN$_{c}$}, where $f_\mathrm{dg}=10^{-N_\mathrm{dg}}$ and $\hat{c}/c=10^{-N_c}$. We define the domain as
$(r,\theta)\in [0.41,95]\,\mathrm{au}\times[\pi/2-0.27,\pi/2+0.27]$ and use a resolution of $N_r\times N_\theta=512 \times 200$, which results in an aspect ratio of $\Delta r/r \Delta \theta \approx 3.25$. This corresponds to $2-8$ cells in the $r$-direction and $6-30$ cells in the $\theta$-direction per scale height $H$ throughout the domain. This resolution is high enough to reproduce the thermal evolution of the self-shadowing features, as demonstrated by the resolution study shown in Appendix \ref{A:ResStudy}.

The domain of these simulations is slightly smaller than in the hydrostatic runs, in such a way that we can simultaneously exclude the zones where the density floor has been applied, where artificial waves are otherwise generated, and impose Dirichlet boundary conditions for all fields by fixing previously calculated values in regions that are now covered by ghost cells. For each $f_\mathrm{dg}$, we use as initial condition the hydrostatic solutions with $t_\mathrm{iter}=100$ yr obtained after $35$ iterations. In each case, we fix at the boundaries the corresponding hydrostatic solutions with $t_\mathrm{iter}=0.1$ yr. We verified that choosing a different boundary condition, for instance using instead the largest-$t_\mathrm{iter}$ hydrostatic solutions at the ghost cells, does not affect the main results of this section, and simply introduces artificial features close to the boundaries which are minimized if the shortest-$t_\mathrm{iter}$ solutions are employed. Our results also hold if the energy is instead set as $E_r=a_R\,(10\,\mathrm{K})^4$ at the ghost cells. To avoid artificial waves continuously forming at the left boundary and at the domain corners, where the density is several orders of magnitude lower than in the rest of the domain, we impose a damping layer in the region $\{r<0.5\,\mathrm{au}\}\cup\{|\theta-\pi/2|>0.12+0.13\, r/\mathrm{au}\}$ (see Fig. \ref{fig:rh_T2D}), in which the irradiation flux is not updated from its boundary condition distribution and all other fields are exponentially damped to that solution with a characteristic time of $0.1/\Omega(R)$. We solve Eqs. \eqref{Eq:RadHD} using third-order Runge-Kutta time integration for the gas fields and its corresponding IMEX version \citep{MelonFuksman2019} for the radiation fields. We compute fluxes for hydrodynamics and radiation fields using the HLLC solvers by \cite{Toro} and \cite{MelonFuksman2019}, respectively. Spatial reconstruction is performed by means of the third-order WENO method by \cite{Yamaleev2009} modified as in \cite{Mignone2014reconstruction} for spherical grids.
We compute the evolution of the system in each case up to at least $500$ yr, using a Courant factor of $0.3$ for the time step computation.

Fig. \ref{fig:rh_T2D} shows a series of time snapshots in all of our simulations with $\hat{c}/c=10^{-3}$ showing the temperature distribution at different points in the disk's evolution. For $f_\mathrm{dg}=10^{-4}$, the temperature remains at all times approximately equal to its initial state, since the employed initial distribution was nearly convergent. Some differences of around $10\%$ with respect to the initial temperature can be seen near the outer radial boundary in Fig. \ref{fig:rh_dT2D}, in which we show for the same snapshots the ratio between current and initial temperature values. This behavior is expected, since the initial state is not exactly at equilibrium. We also observe in this run a transient formation of sound waves at the boundaries due to the initial relaxation of the system, which we further describe later in this section.

\begin{figure*}[t!]
\centering
\includegraphics[width=\linewidth]{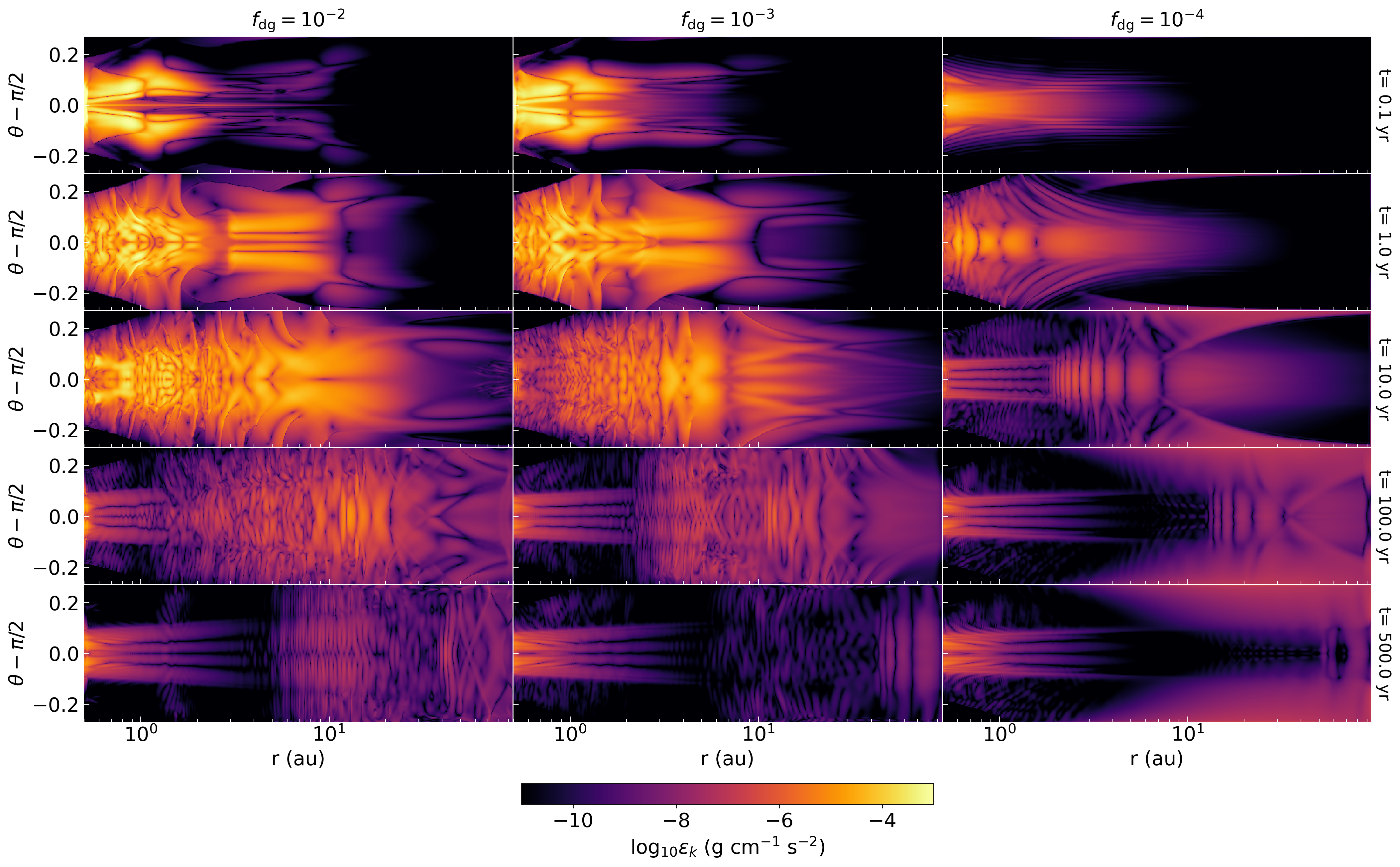}
\caption{Same snapshots as Fig. \ref{fig:rh_dT2D}, showing the time evolution of $\log_{10}\epsilon_k$, where $\epsilon_k=\frac{1}{2}\rho(v_r^2+v_\theta^2)$.}
\label{fig:rh_ekin}
\end{figure*}

Much more relevant is the behavior in \sftw{rhdg2\_c3} and \sftw{rhdg3\_c3}, which have shadowed regions at $t=0$. As soon as simulations are started, even before sound waves are produced at the boundary, we observe an immediate thermal relaxation of the outer, optically thin disk layers. Fig. \ref{fig:rh_dT2D} shows that overheated regions start to cool down and shadowed regions heat up, which blurs the initial temperature perturbations at the disk surface. This causes an immediate reconfiguration of the radiation flux emitted at the overheated regions, which redirects itself towards the previously shadowed ones. This can be seen in Fig. \ref{fig:rh_f}, where we show the evolution of the reduced flux and the flux direction. The same figure shows that the vertical temperature gradient produced by the reconfiguration of the outer layers leads to diffusion towards the midplane in shadowed regions and away from it in overheated ones. At approximately $0.1$ yr, outward-traveling sound waves start being emitted at the inner radial boundary due to the system's initial relaxation, becoming shocks at the disk surface (see Fig. \ref{fig:rh_ekin}). These waves carry more energy as in the $f_\mathrm{dg}=10^{-4}$ runs and are even visible in the temperature distributions, since solutions are farther apart from the final equilibrium configuration as in that case.

No signs of propagation of the temperature bumps towards the star are observed either before or after the formation of the relaxation-induced sound waves. Instead, outward-traveling waves keep forming as the thermally relaxed regions grow towards the midplane. The relaxed layers reach the midplane faster for larger radii (see Fig. \ref{fig:rh_dT2D}), due to the lower vertical optical depth an consequently shorter diffusion times away from the star. All radial temperature extrema are eventually erased after approximately $100$ and $10$ yr for runs \sftw{rhdg2\_c3} and \sftw{rhdg3\_c3} respectively (see Fig. \ref{fig:rh_T2D}). These times coincide in each case with the relaxation timescale $t_\mathrm{cool}$ estimated in Section \ref{SS:RelaxationTime} for surface temperature perturbations reaching the midplane.
%After that time, the sound waves stop being produced as the system has reached hydrostatic equilibrium, and the perturbations induced by them on the temperature and flux distributions eventually disappear (see Fig. \ref{fig:rh_f}).
After $500$ yr in \sftw{rhdg2\_c3} and $100$ yr in \sftw{rhdg3\_c3}, the temperature distributions look like the hydrostatic solutions obtained for $t_\mathrm{iter}=0.1$ yr and no hints of formation of new temperature maxima can be seen.

\begin{figure*}[t!]
\centering
\includegraphics[width=\linewidth]{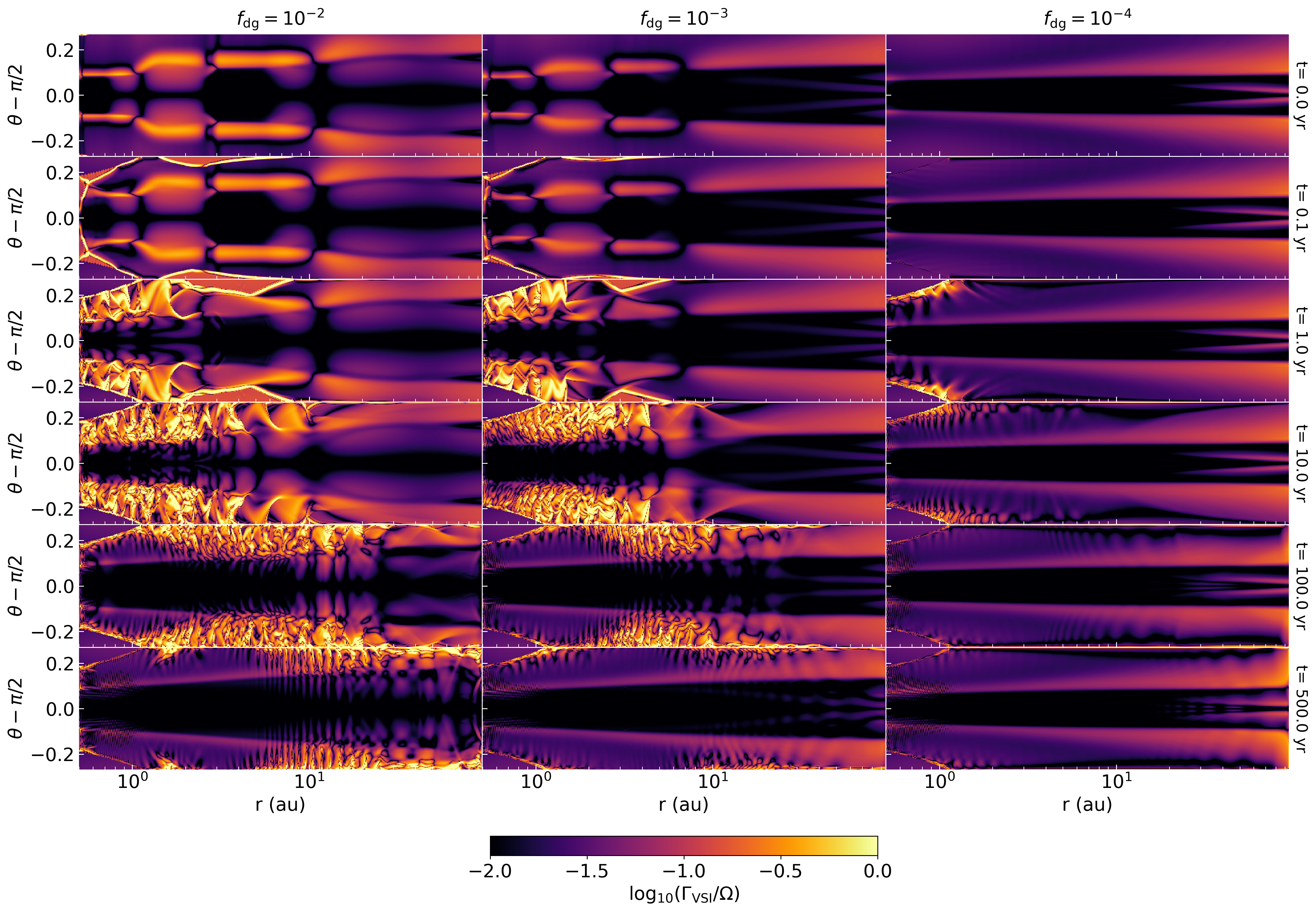}
\caption{Same as Fig. \ref{fig:rh_T2D}, showing %$\log_{10}(\Gamma_\mathrm{VSI}/\Omega)$, where $\Gamma_\mathrm{VSI}$ is 
the expected linear growth rate of the fastest-growing VSI mode for instant thermal relaxation normalized by the local $\Omega$.}
\label{fig:rh_VSI}
\end{figure*}

We now take a closer look at the gas motion during the system's initial relaxation, focusing both on the damping of the initial temperature extrema and on the disk's hydrodynamical stability. We highlight the regions of space where the kinetic energy is concentrated by computing its component due to meridional motion, given by $\epsilon_k=\frac{1}{2}\rho(v_r^2+v_\theta^2)$. Snapshots of this quantity are shown in Fig. \ref{fig:rh_ekin}, where we can see that the thermal relaxation seen in Fig. \ref{fig:rh_dT2D} is accompanied by a reconfiguration of the gas structure via advection in the entire disk before this is crossed by the relaxation sound waves. We also see that these waves stop being emitted as soon as the disk reaches its final equilibrium configuration. After $t\sim t_\mathrm{cool}$, the remaining waves are slowly dispersed as they travel outwards. 
% Additional vertical compression and expansion-like features can be seen close to the inner radial boundary, caused by the constant forcing at the damping layer. Slower, lower-amplitude waves excited at the damping zone, possibly buoyancy or internal gravity modes \citep[see, e.g.,][]{Lubow1998,vandenBremer2014}, can also be seen in these snapshots after $500$ yr.
The additional boundary- and damping-induced features occurring close to the inner radial boundary at larger times are of no relevance for the conclusions of this paper, as they do not affect the overall disk structure.

During the disk relaxation, the gas motion shows velocity fluctuations at the disk atmosphere. Some of these are caused by the evolution of the disk towards a new equilibrium configuration, but these could also be signposts for an evolving hydrodynamical instability, as it occurs at much higher resolution, e.g., in the 3D studies by \cite{Flock2017VSI} of the VSI in optically thin regions ($>20$ au) of irradiated disks. We have explored this possibility by first verifying that the disk is always radially Rayleigh-stable in the entire domain, since the squared epicyclic frequency $\kappa_R^2$ for radial perturbations, computed as $\kappa_R^2=\frac{1}{R^3}\partial_R(R^4 \Omega^2)$, is always positive. However, due to its nontrivial radial and vertical temperature stratification, the disk can still be COS- and VSI-unstable at various levels in the entire domain (Klahr et al. 2022, submitted).  Assuming either instant relaxation at the atmosphere or that radiative diffusion defines an optimally cooling wavenumber \citep{Latter2018}, we can get an idea of the order of magnitude of the local expected VSI growth rate $\Gamma_\mathrm{VSI}$ using the expression for the fastest-growing local VSI mode for instant relaxation given by $\Gamma_\mathrm{VSI}=\frac{|\kappa_z^2|}{2\Omega}$ (Klahr et al. 2022, submitted), where the squared vertical epicyclic frequency $\kappa_z^2$ can be obtained as $\kappa_z^2=\frac{1}{R^3}\partial_z(R^4 \Omega^2)$.
% In this case, the sign of the squared epicyclic frequency does not provide a criterion for stability, and just determines the direction of the most unstable wave vector.
We show in Fig. \ref{fig:rh_VSI} snapshots of $\Gamma_\mathrm{VSI}$ normalized by the local $\Omega$.
% At $t=0$, the pressure and density stratification in both the radial and vertical directions induced by the temperature bumps leads to enlarged growth rates close to the disk surface, which reach $\Gamma_\mathrm{VSI}/\Omega \sim 0.5$ and $0.2$ in \sftw{rhdg2\_c3} and \sftw{rhdg3\_c3}, respectively. However, no VSI growth is seen in those regions at $t=0.1$ yr. At later times, the transient stratification caused by the crossing waves leads to higher linear growth rates at the disk atmosphere, reaching $\Gamma_\mathrm{VSI}/\Omega \sim 5$ in both \sftw{rhdg2\_c3} and \sftw{rhdg3\_c3}. It is then possible that the observed velocity fluctuations are linked to the growth of the VSI in such regions. In \sftw{rhdg4\_c3}, large growth rates of up to $\Gamma_\mathrm{VSI}/\Omega \sim 1$ only occur at the edge of the damping region, and no strong fluctuations are observed.
In \sftw{rhdg2\_c3} and \sftw{rhdg3\_c3}, the transient stratification caused by the crossing waves leads to high growth rates at the atmosphere ($\Gamma_\mathrm{VSI}/\Omega\sim 5$), which makes it possible that the observed velocity fluctuations are linked to VSI growth. In \sftw{rhdg4\_c3}, the growth rates are smaller and no strong fluctuations are observed.
In all cases, regions of large growth rates vanish at large times, and only boundary- and damping-induced features remain. The resulting stability is most likely caused by the fact that our employed radial resolution of $2-8$ cells per scale height does not suffice to obtain spontaneous VSI growth arising from a hydrostatic stratification within our simulated run times \citep[see, e.g., the resolution study in][]{Flores2020}. Especially since a diffusive cooling process occurs in the optically thicker parts, the necessary small wavelengths that could still show significant growth are not resolved \citep[][]{Latter2018}. Higher-resolution studies of the VSI employing the same techniques as in this work, currently in preparation, will be the matter of a next article.

A conclusion regarding the disk's cooling can be drawn from the fact that the vertical extent of the thermally relaxed layers in Figs. \ref{fig:rh_T2D} and \ref{fig:rh_dT2D} does not trace the relaxation-induced waves, which suggests that it is unlikely that these and the mentioned velocity fluctuations play an important role in the vertical transport of energy. Such process seems to be instead entirely dominated by the joint effect of radiative cooling and gas advection, which starts smoothing out temperature extrema even before waves are formed. It is possible, however, that radial transport of energy due to either dissipation of kinetic energy or advection is enhanced by the radially travelling waves.

\subsection{Reduced speed of light approximation}\label{SS:RSLA}

\begin{figure}[t!]
\centering
\includegraphics[width=\linewidth]{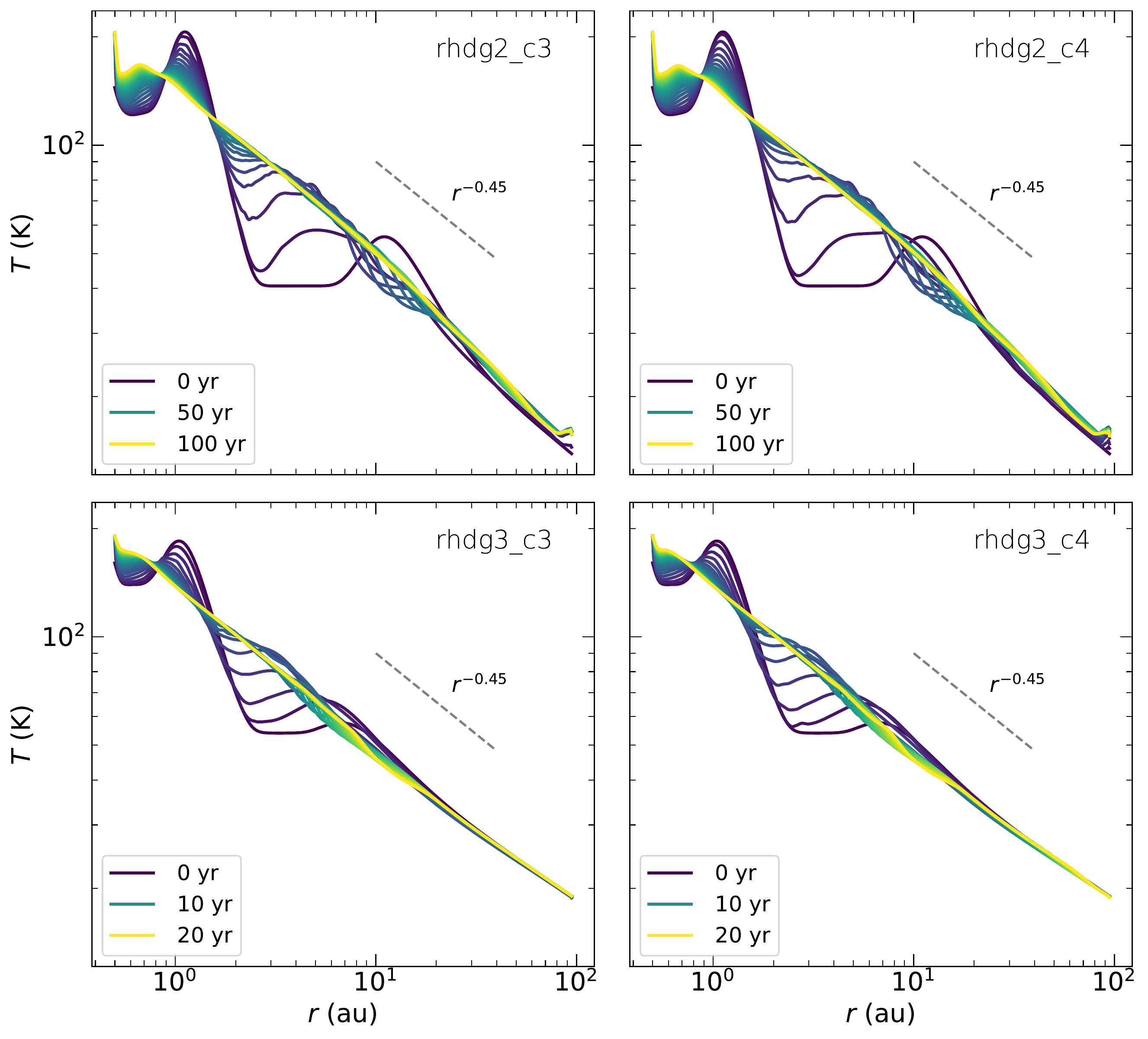}
\caption{Midplane temperature as a function of radius measured at different times for $f_\mathrm{dg}=10^{-2}$ (top) and $10^{-3}$ (bottom), taking $\hat{c}/c=10^{-3}$ (left) and $10^{-4}$ (right). The power law $r^{-0.45}$ used at the first iteration for the construction of initial conditions is shown for comparison (dashed lines).}
\label{fig:rh_Tmidplane}
\end{figure}

\begin{figure}[t!]
\centering
\includegraphics[width=\linewidth]{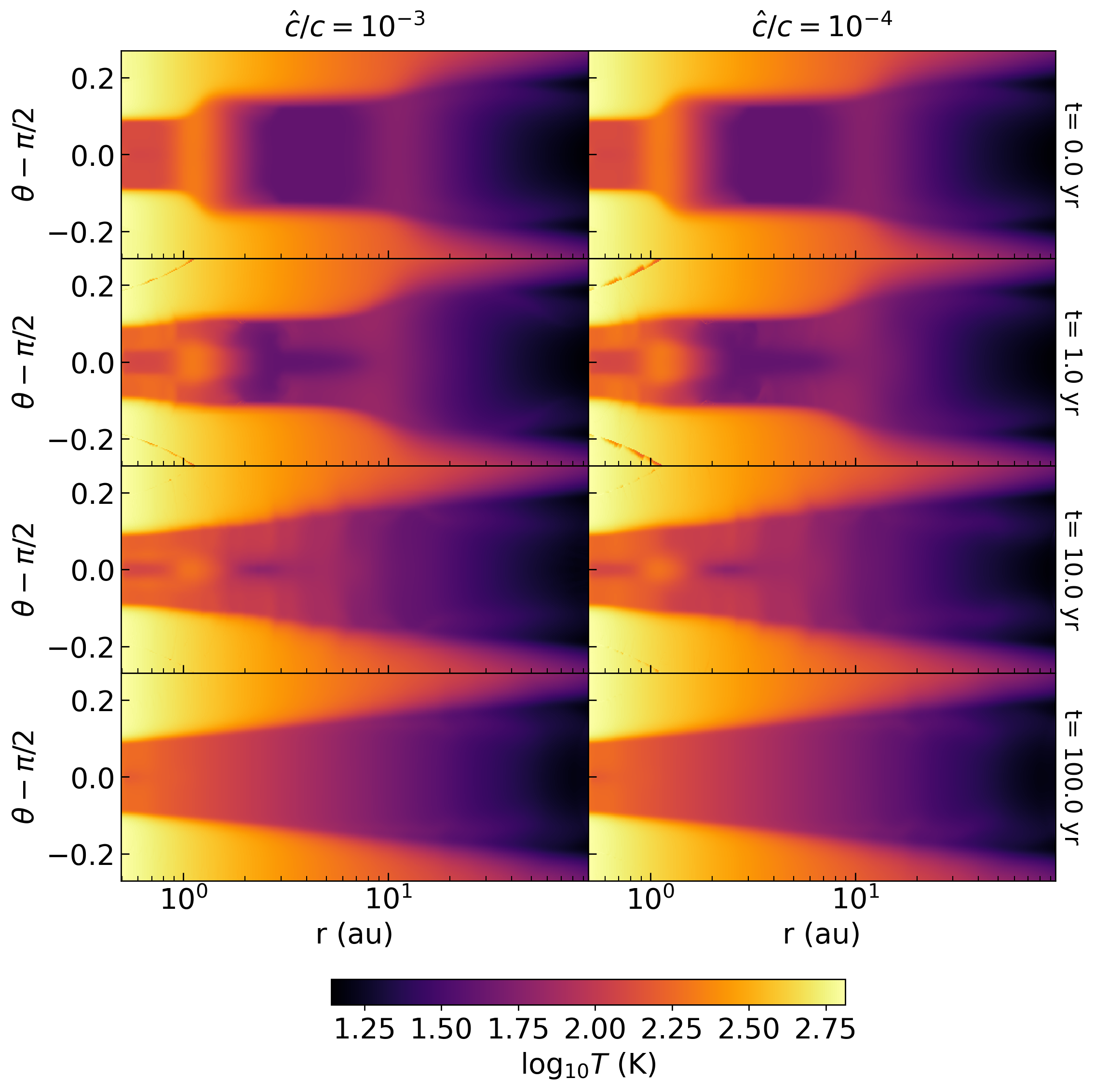}
\caption{Temperature distributions at different times for runs \sftw{rhdg2\_c3} (left) and \sftw{rhdg2\_c4} (right).}
\label{fig:rh_T2D_c}
\end{figure}

\begin{figure}[t!]
\centering
\includegraphics[width=\linewidth]{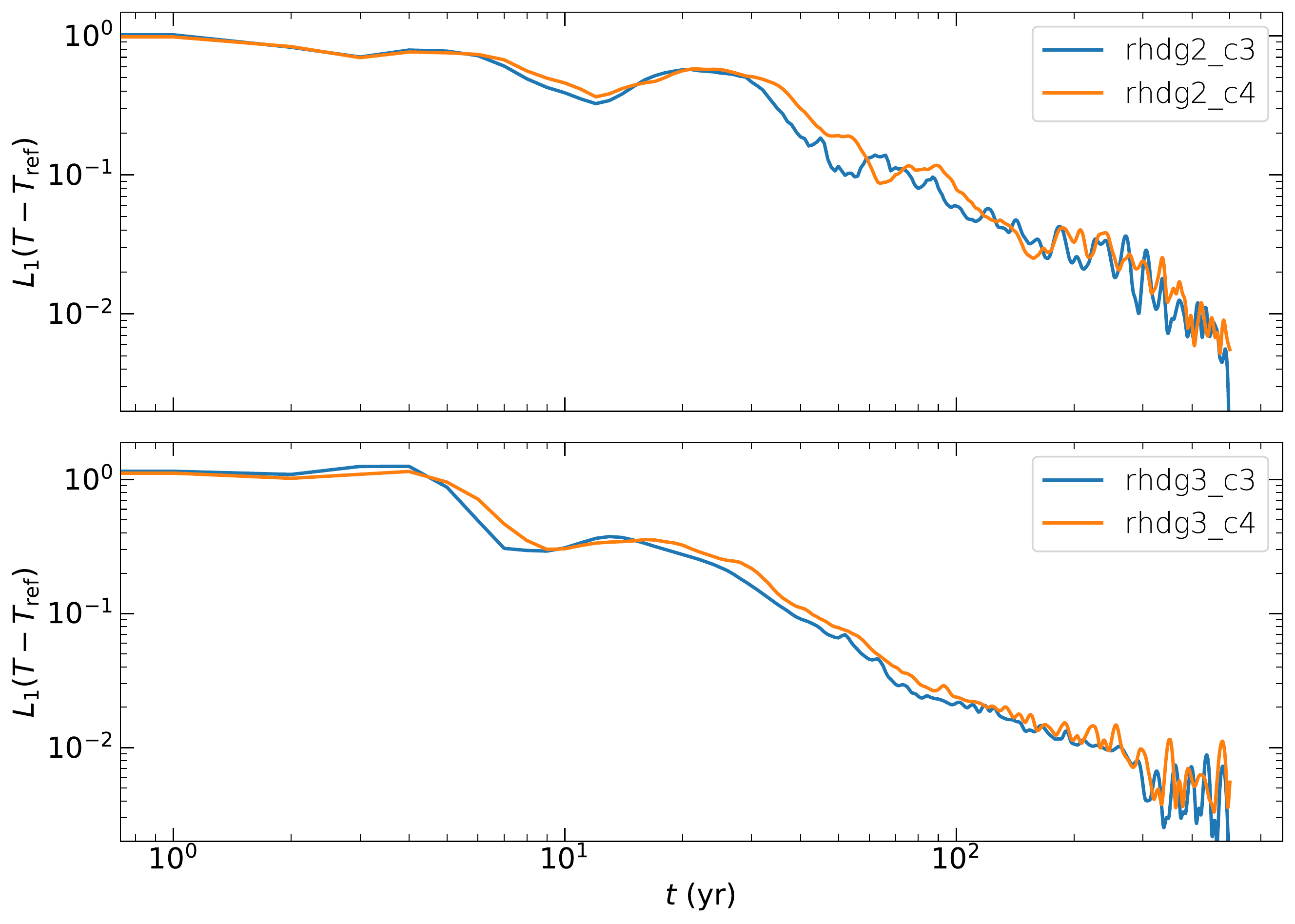}
\caption{Normalized time-dependent $L_1$-norms of $T-T_\mathrm{ref}$ for different values of $\hat{c}$ taking $f_\mathrm{dg}=10^{-2}$ (top) and $10^{-3}$ (bottom).}
\label{fig:rh_L1T}
\end{figure}

An issue that must be carefully treated is if the disk evolution is affected by our choice of of $\hat{c}$. A criterion for the validity of the reduced speed of light approximation was derived in \cite{Skinner2013}, who suggested that $\hat{c}$ must always be larger than the maximum fluid velocity $v_\mathrm{max}$ and that radiation diffusion timescales must always be smaller than typical dynamical timescales. This condition can be summarized as
\begin{equation}\label{Eq:RedCCriterion}
    \hat{c}\gg v_\mathrm{max}\,\max(1,\tau_\mathrm{L})\,,
\end{equation}
where $\tau_\mathrm{L}$ is the optical depth along a characteristic length of the system. If we take $v_\mathrm{max}\sim c_s$, i.e., the gas sound speed, we get that for $\hat{c}/c=10^{-3}$ the ratio $\hat{c}/c_s$ is always of the order of $10^2-10^3$, so the first condition is fulfilled. However, it is harder to define a correct characteristic length for a computation of $\tau_\mathrm{L}$ that is relevant for this particular problem, even more considering that different radiative transport regimes dominate in different disk regions. Nevertheless, Eq. \eqref{Eq:RedCCriterion} is an approximate relation, and the validity of this approach can only be safely assessed through careful testing of the problem at hand \cite[see, e.g.,][]{MelonFuksman2021}. We therefore follow the approach of Section \ref{SS:RelaxationTime} and directly compare results using different values of $\hat{c}$.

Fig. \ref{fig:rh_Tmidplane} shows a time series of the midplane temperature as a function of radius for all simulations with $f_\mathrm{dg}=10^{-2}$ and $10^{-3}$, computed with $\hat{c}/c=10^{-3}$ and $10^{-4}$. For each $f_\mathrm{dg}$, the functional form of these profiles is roughly the same at all times independently of $\hat{c}$, with slight differences in the perturbations caused by the sound waves. At $t\sim t_\mathrm{cool}$ for all $\hat{c}$, the midplane temperature reaches a similar power-law functional dependence with radius as that used at the beginning of the iterative procedure in Section \ref{S:DiskModels} ($T\propto r^{-0.45}$), with some deviations at the radial boundaries. We have also compared the full 2D temperature distributions at different times for different values of $\hat{c}$, shown for $f_\mathrm{dg}=10^{-2}$ in Fig. \ref{fig:rh_T2D_c}. These profiles look approximately the same independently of $\hat{c}$, again with slight differences induced by the waves. Both simulations have at all times thermally relaxed surface layers of the same vertical extent, and reach their final configuration at the same time. As in Section \ref{SS:RelaxationTime}, this happens because, provided the adjustment of the radiation fields is rapid enough, gas cooling only depends on the $c\, G^0$ term in Eq. \eqref{Eq:RadHD}. Lastly, we have made a space-averaged comparison of the $f_\mathrm{dg}=10^{-2}$ and $10^{-3}$ runs for all times by computing the $L_1$ norm of $T-T_\mathrm{ref}$, where, for each $f_\mathrm{dg}$, $T_\mathrm{ref}$ is the temperature in the $t=500$ yr snapshot of the corresponding run with $\hat{c}/c=10^{-3}$. We show in Fig. \ref{fig:rh_L1T} the values of $L_1(T(t)-T_\mathrm{ref})$ as a function of time normalized by $L_1(T(0)-T_\mathrm{ref})$. The obtained curves are almost identical for both $\hat{c}$ values in all cases. We conclude that the observed suppression of temperature extrema in all cases is not affected by our choice of $\hat{c}$.

\subsection{Discussion}\label{SS:RadHDiscussion}

We saw in Section \ref{S:DiskModels} that a self-shadowing instability would require surface temperature perturbations to reach the midplane faster than the gas adjusts to thermal changes. On the contrary, the results of this section show that temperature extrema are smoothed out at the surface long before the temperature is altered at the midplane, and therefore an instability producing temperature perturbations throughout the full vertical extent of the disk as in \sftw{dg2t100} and \sftw{dg3t100} can never occur. Following this argument, the only remaining possibility would be that such an instability occurred in outer disk layers that are thin enough to transmit temperature perturbations in the vertical direction before the gas readjusts its structure, causing local increases in the scale height similar to those seen, e.g., in the hydrostatic run \sftw{dg2t1}. However, this process is likely disallowed by radial radiative transfer, as such layers would simultaneously have to be optically thin enough to allow for fast vertical diffusion and optically thick enough to prevent radiative transport from damping temperature extrema in the radial direction. As discussed in Section \ref{SS:RelaxationTime}, this is supported by the fact that no such outer layers are seen for short enough $t_\mathrm{iter}$. Therefore, our results support the hypothesis that circumstellar disks are stable to self-shadowing due to the fact that surface temperature perturbations are damped by radiative cooling and gas advection before they affect lower, optically thick regions via diffusion. This also justifies the use of simplified power-law prescriptions for the midplane temperature in 1D disk models. Furthermore, our results imply that non-convergent hydrostatic disk models obtained through iterative procedures may not reflect the actual disk structure if they underestimate the vertical diffusion time, as it is the case if the temperature is computed by means of Monte Carlo methods.

It remains to be seen if self-shadowing effects can occur if self-sustained scale height perturbations are induced by external factors, such as planetary perturbers, large-scale structures formed by hydrodynamical instabilities, or complex dust distributions inducing local changes in the opacity. In particular, opacity transitions at different ice lines can cause sharp, sustained temperature variations that could potentially evolve through self-shadowing. We intend to study this process in future investigations. Future studies will also have to account for the damping of long-standing temperature perturbations by turbulent mixing and viscosity, which we have not studied in this work. On the other hand, irradiation-induced structures unrelated to self-shadowing can still occur in disks around variable sources such as EX Lupi and FU Orionis stars, in which sudden outbursts can excite outward-traveling sound and gravity waves \citep{Schneider2018}.

Our assumption of a constant $f_\mathrm{dg}$ does not account for vertical settling and depletion of dust in the disk atmosphere, which could in principle reduce the height of the different $\tau=1$ surfaces shown in Fig \ref{fig:2Dtausurfaces}. Even though the Stokes number remains smaller than $1$ for all considered particle sizes well above the $\tau=1$ surfaces for irradiation, the actual dust distribution in the atmosphere depends on the balance between turbulent stirring and vertical settling. The dust distribution away from the midplane is better captured by hydrodynamical simulations than by simple prescriptions relying on a parameterization of the turbulence strength \citep[see,e.g.,][]{Dubrulle1995,DullemondDominik2004dust}, since it depends on the particular source of turbulence. An example of this is the anisotropic turbulence produced by the VSI, whose vertically elongated modes significantly increase the vertical stirring of dust with respect to models of homogeneous isotropic turbulence \citep{Stoll2016,Stoll2017}. We note that these authors obtain a vertical Gaussian dust distribution of micron-sized particles with the same scale height as the gas in the entirety of a domain reaching $z=5 H$ at $5$ au, which supports the validity of our prescription. On the other hand, dust sedimentation of small grains in outer regions can even be affected by non-ideal magnetohydrodynamical effects and magnetized winds \citep{Riols2018,Hutchison2021,Booth2021}. In principle, we do not expect dust settling effects to be relevant for our work as long as they occur in optically thin regions where radiative transport prevents any shadowing from occurring. Future disk models could test this hypothesis by studying the stability of disks in which the opacity is computed as a function of a dynamically evolving small dust distribution. 
We have also assumed instantaneous thermal equilibrium of gas and dust particles, which is not verified at high altitudes as collisions between dust and gas particles become less frequent for smaller densities \citep[see, e.g.][]{Malygin2017,Pfeil2021}. We do not expect this effect to alter our results, since at most it could increase the time it takes to form temperature perturbations in such regions. 
 \section{Conclusions}\label{S:Conclusions}
 
 We conducted the first 2D radiation-hydrodynamical simulations exploring the so-called self-shadowing instability in protoplanetary disks, using the two-moment Rad-HD module by \cite{MelonFuksman2021} integrated in the finite-volume \sftw{PLUTO} code. Our model includes frequency-dependent stellar irradiation, and employs typical star and disk parameters for a T Tauri system. All opacities were computed assuming a uniform dust-to-gas mass ratio $f_\mathrm{dg}$ for sub-micron dust grains. We explored the amplification of thermal perturbations in hydrostatic models for different $f_\mathrm{dg}$ values and their thermal relaxation when advection is switched back on. We summarize our main conclusions as follows:
 
 \begin{itemize}
     \item Our iterative procedure for the construction of hydrostatic configurations does not converge for high enough $f_\mathrm{dg}$ and long enough time $t_\mathrm{iter}$ used for the thermal evolution between consecutive iterations. In such cases, inward-traveling vertically isothermal temperature perturbations are formed in the entire optically thick region of the disk due to self-shadowing. These results are consistent with other works employing similar iterative methods. 
     \item When $t_\mathrm{iter}$ is shorter than the vertical diffusion time, the temperature perturbations produced at the irradiation surface do not reach the midplane before a new iteration is started. This effect cannot be captured by models that assume instantaneous radiative diffusion. As a consequence, a middle layer is formed in which temperature variations are smaller than at the surface.  The thickness of the middle layer increases for decreasing $t_\mathrm{iter}$ and scale height perturbations are not formed for small enough $t_\mathrm{iter}$, which shows that a self-shadowing instability is only possible if thermal perturbations reach the optically thick disk region faster than the disk evolves at the surface layers.
     \item We ran Rad-HD simulations taking as initial conditions the hydrostatic solutions constructed with the highest $t_\mathrm{iter}$ values once temperature bumps are fully formed. As soon as simulations start, these perturbations begin to be damped in the outer disk layers through radiative transport and gas advection. The thermally relaxed layers grow towards the midplane until a vertically isothermal hydrostatic configuration is reached after a vertical diffusion time of about $100$ yr for $f_\mathrm{dg}=10^{-2}$ and $10$ yr for $f_\mathrm{dg}=10^{-3}$.
     Radial temperature maxima are entirely smoothed out close to the surface before the thermally relaxed layers reach the midplane, which shows that surface relaxation is much faster than vertical diffusion.
     \item Outward-traveling waves are formed during the system relaxation without substantially intervening in the vertical transport of energy. During this transient phase the gas shows velocity fluctuations at the atmosphere, which are possibly linked to the triggering of VSI by the density and pressure stratification produced by the waves. However, no VSI is seen once the system relaxes, due to the low employed resolution in the radial direction. 
     \item None of our results depend on our chosen value of the reduced speed of light $\hat{c}$, since after the radiation field quickly reaches a stationary configuration, gas cooling is entirely regulated by the source term in the gas energy equation, which uses the physical value of $c$. We verified this by comparing the results of the same Rad-HD simulations run with different $\hat{c}$ values.
 \end{itemize}
 
 Our findings suggest that circumstellar disks are stable to self-shadowing due to the damping of temperature perturbations at the surface through radiative cooling and advection before these affect optically thick disk regions via diffusion. This also justifies the use of disk models where the midplane temperature scales with the distance to the star following a simple power law. Furthermore, we showed that iterative methods used to obtain hydrostatic models of irradiated disks may induce a formation of scale height perturbations if they underestimate the vertical diffusion time, as it is the case for non-convergent Monte Carlo-based models used in previous works. Further work is needed to test our results if the assumption of a constant dust-to-gas mass ratio is replaced by a more realistic model for the evolution of dust in a turbulent disk.

\appendix
 
 \section{Optical depth computation}\label{A:OptDepth}

In the hydrostatic runs in Section \ref{S:DiskModels}, we approximate $\tau_0$ in Eq. \eqref{Eq:OpticalDepth} by assuming that the part of the disk left outside of the domain occupies a radial extent $r\in[r_0,r_\mathrm{min}]$. We assume in that region a midplane temperature $T\propto R^{q}$ and a midplane density $\rho(r,\pi/2)\propto R^p$, where $p$ and $q$ are the same exponents used to compute the density and temperature profiles at the first iteration. The gas density distribution for a vertically isothermal disk with these parameters is
\begin{equation}\label{Eq:RhoIsothermal}
    \rho(R,z) = \rho_0 
    R^p
    \exp \left[
    \frac{R^2}{H^2}
    \left(
    \frac{R}{\sqrt{R^2+z^2}}-1
    \right)
    \right]\,,
\end{equation}
which in spherical coordinates becomes
\begin{equation}
    \rho(r,\theta) = \rho_0\, r^p
    \sin^p\theta
    \exp \left[
    \frac{\sin\theta-1}{
    H_0^2\, r^{q+1}\, \sin^{q+1}\theta
    }
    \right]\,,
\end{equation}
where $H=H_0\, R^{(q+3)/2}$ and all lengths are normalized by $1$ au. Evaluating this expression at
$(r_\mathrm{min},\theta)$ and $(r_\mathrm{min},\pi/2)$, we obtain:
\begin{equation}
\begin{split}
    \rho(r_\mathrm{min},\theta) &= \rho_0\, r_\mathrm{min}^p
    \sin^p\theta
    \exp \left[
    \frac{\sin\theta-1}{
    H_0^2\, r_\mathrm{min}^{q+1}\, \sin^{q+1}\theta
    }
    \right] \\
    \rho(r_\mathrm{min},\pi/2) &= 
    \rho_0\, r_\mathrm{min}^p \,.
\end{split}
\end{equation}
Using these equations, the density can be rewritten using only its value at $r_\mathrm{min}$:
\begin{equation}\label{Eq:rho_pq}
  \rho(r,\theta) = 
  \left(
  \frac{\rho(r_\mathrm{min},\theta)}{\rho(r_\mathrm{min},\pi/2)
  \sin^p\theta}
  \right)^{\left( \frac{r_\mathrm{min}}{r} \right)^{q+1}}
  \rho(r_\mathrm{min},\pi/2)
  \left(\frac{r}{r_\mathrm{min}}\right)^p
  \sin^p\theta \,.
\end{equation}
We use this expression to estimate the density in the entire region $r\in[r_0,r_\mathrm{min}]$, and compute $\tau_0$ for all frequencies as
\begin{equation}\label{Eq:Tau0}
    \tau_0(r_\mathrm{min},\theta,\nu)
    = \kappa_\nu\,
    \int^{r_\mathrm{min}}_{r_0}
    \mathrm{d}r'\,
    \rho_\mathrm{d}(r',\theta)\,,
\end{equation}
where $\rho_d=f_\mathrm{dg}\,\rho$. We compute this integral analytically by defining
a radial grid in the region $r\in[r_0,r_\mathrm{min}]$. On the other hand, in the Rad-HD simulations, $\tau_0$ is obtained by adding its already computed value in the hydrostatic runs to the optical depth of the new left-out portion of the disk when the domain is reduced. The resulting $\tau_0(r_\mathrm{min},\theta,\nu)$ is assumed to remain constant in time, and therefore this computation is only carried out at the beginning of each run.

\begin{figure}[t!]
\centering
\includegraphics[width=0.7\linewidth]{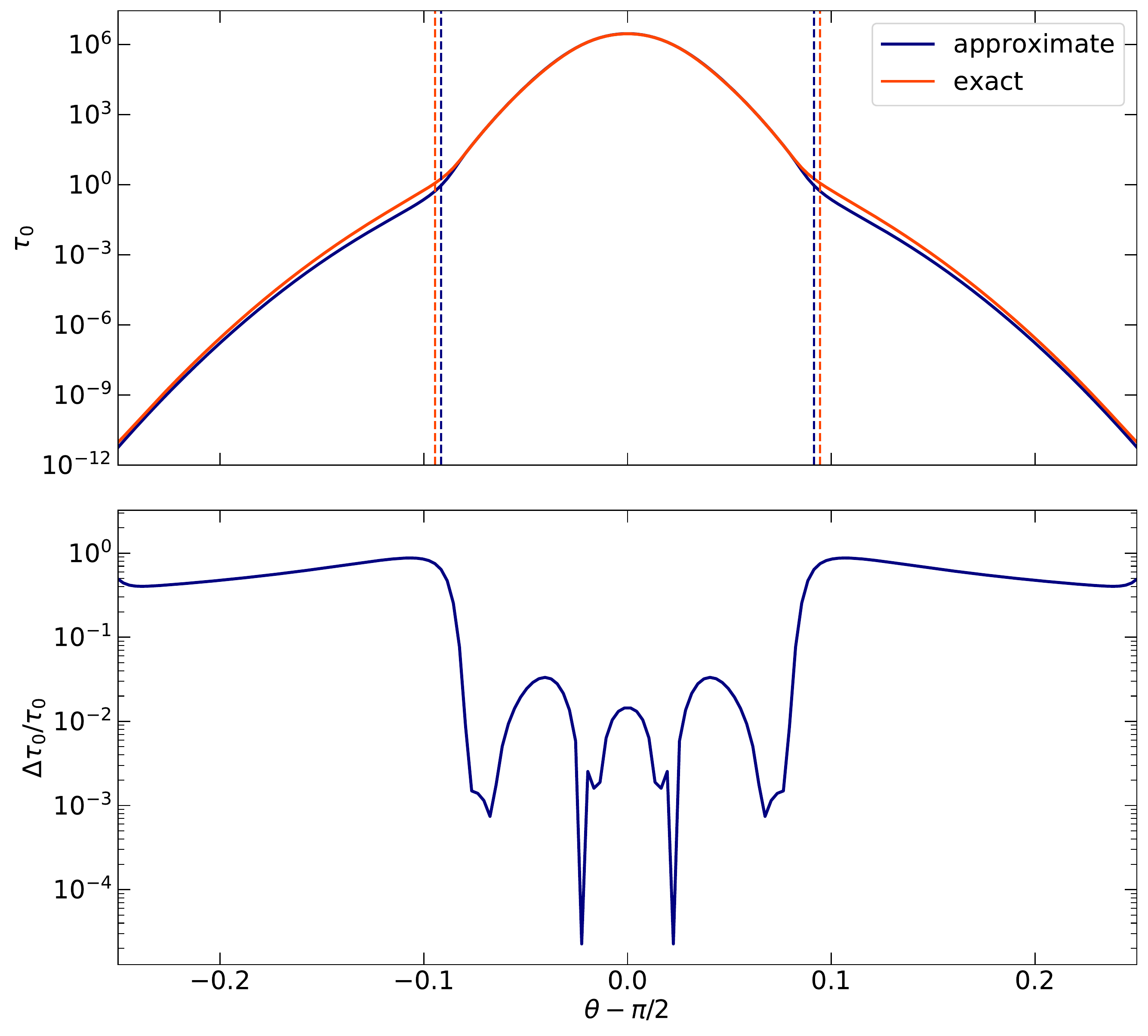}
\caption{Top: exact and approximate (using Eq. \eqref{Eq:rho_pq}) $\tau_0(\theta)$ profiles computed between $r=0.1$ and $0.4$ au in a disk with a vertical temperature stratification. Vertical dashed lines indicate the location of the $\tau=1$ surfaces for each method. Bottom: relative $\tau_0$ differences between both approaches.}
\label{fig:tau0}
\end{figure}

Eqs. \eqref{Eq:rho_pq} and \eqref{Eq:Tau0} yield a better prescription for $\tau_0$ than, e.g., $\tau_0(r_\mathrm{min},\theta,\nu)= \kappa_\nu\rho(r_\mathrm{min},\theta)(r_\mathrm{min}-r_0)$ \citep{Flock2013}, since the resulting $\tau_0$ is exact for vertically isothermal disks. However, real disks have a sharp transition to the atmosphere’s temperature at a few scale heights above the midplane (Figs. \ref{fig:Ttheta}, \ref{fig:2Dtausurfaces}), where the $\tau=1$ surfaces for irradiation are located. We tested the accuracy of our prescription in such cases by using Eq. \eqref{Eq:rho_pq} to compute $\tau_0$ in a chosen radial range for a density distribution with a vertical temperature gradient. We used as test distribution a hydrostatic configuration obtained as in Section \ref{S:DiskModels} with $f_\mathrm{dg}=10^{-2}$ and $r_\mathrm{min}=0.1$ au, i.e., with a minimum radius close to to $r_0$. We computed the optical depth in this solution between $0.1$ and $0.4$ au, which would correspond to $\tau_0$ in a simulation with $r_\mathrm{min}=0.4$ au. We also performed the same calculation taking $\rho$ from Eq. \eqref{Eq:rho_pq} with $r_\mathrm{min}=0.4$ au. Unlike for the computation of $\mathbf{F}_\mathrm{Irr}$, we used in both cases $\kappa_\nu=\kappa_P(T_s)$. This allows us to estimate the location of the irradiation surface by equating $\tau_0=1$ and to evaluate the error of the prescription, which only depends on the integral of $\rho_\mathrm{d}$. We show a comparison of both $\tau_0(\theta)$ curves and their relative difference in Fig. \ref{fig:tau0}. We see that our prescription leads to relative $\tau_0$ differences close to $1$ at the $\tau=1$ surfaces. However, the value of $\tau_0$ decreases so sharply with altitude that this barely changes the location of the $\tau=1$ surface an amount $\Delta\theta\approx 0.003$ coinciding with our grid resolution, since both $\tau_0$ curves are almost identical. This justifies the computation of $\tau_0$ using Eq. \eqref{Eq:rho_pq} even for disks with a vertical temperature gradient.

 \section{Implicit method}\label{A:ImplicitMethod}

In all IMEX schemes in the Rad-HD module, radiation-matter source terms are implicitly integrated by solving an equation system of the form
\begin{equation}\label{Eq:ImplEq}
    \begin{pmatrix}
    E_r \\
    \mathbf{F}_r
    \end{pmatrix}
    = \begin{pmatrix}
    E_r \\
    \mathbf{F}_r
    \end{pmatrix}'
    - s \hat{c} \Delta t
    \begin{pmatrix}
    G^0 \\
    \mathbf{G}
    \end{pmatrix}\,,
\end{equation}
together with the modified conservation equation
\begin{equation}\label{Eq:EFConsEqs}
    \begin{pmatrix}
    E + \frac{c}{\hat{c}}E_r \\
    \rho \mathbf{v} + \frac{1}{\hat{c}}\mathbf{F}_r
    \end{pmatrix}
    =
    \begin{pmatrix}
    E_\mathrm{tot} \\
    \mathbf{m}_\mathrm{tot}
    \end{pmatrix}
    =
    \begin{pmatrix}
    E + \frac{c}{\hat{c}}E_r \\
    \rho \mathbf{v} + \frac{1}{\hat{c}}\mathbf{F}_r
    \end{pmatrix}'\,,
\end{equation}
where $s$ is a constant, $\Delta t$ is the current time step, and primed quantities are computed at the previous step. The full form of the source terms in the laboratory frame is
\begin{equation}\label{Eq:SourceTerms}
  \begin{split}
  G^0 &= \rho \kappa \left( E_r - a_R T^4 -
  2 \bm\beta \cdot \mathbf{F}_r \right)
  +\rho\chi\,\bm\beta\cdot\left(
		\mathbf{F}_r-E_r\bm\beta-\bm\beta \cdot\mathbb{P}_r  
  \right)\\
  \mathbf{G} &= \rho \kappa \left( E_r - a_R T^4 -
  2 \bm\beta \cdot \mathbf{F}_r \right)\bm\beta 
  +\rho\chi\bm\left(
		\mathbf{F}_r-E_r\bm\beta-\bm\beta \cdot\mathbb{P}_r  
  \right)\,,
  \end{split}
  \end{equation}
where $\bm\beta=\mathbf{v}/c$ and $\kappa$ and $\chi$ are the absorption and total opacity coefficients, computed as general functions of the Rad-HD fields (in this work, $\kappa=f_\mathrm{dg} \kappa_P (T)$ and $\chi=f_\mathrm{dg}\kappa_R (T)$). In \cite{MelonFuksman2021}, Eq. \eqref{Eq:ImplEq} is implicitly solved via a four-dimensional root-finding algorithm, either fixed-point or Newton's method. Here we have implemented a faster method similar to the implicit scheme in \cite{Skinner2013}, in which all terms of order $\bm\beta$ and $\bm\beta^2$ are explicitly integrated during the explicit step of the IMEX scheme, as they are small enough to stably do so. The remaining equation to be implicitly solved is
\begin{equation}\label{Eq:ImplEq2}
    \begin{pmatrix}
    E_r \\
    \mathbf{F}_r
    \end{pmatrix}
    = \begin{pmatrix}
    E_r \\
    \mathbf{F}_r
    \end{pmatrix}'
    - s \hat{c} \Delta t
    \begin{pmatrix}
    \rho \kappa (E_r-a_R T^4) \\
    \rho \chi \mathbf{F}_r
    \end{pmatrix}\,,
\end{equation}
 together with Eq. \eqref{Eq:EFConsEqs}. The first of these equations can be written as a polynomial in the specific internal energy $\epsilon$ (Eq. \eqref{Eq:EoS}) as
 \begin{equation}\label{Eq:ImplPol}
     A \epsilon^4 + B \epsilon + C = 0\,,
 \end{equation}
 with
 \begin{equation}\label{Eq:ImplABC}
    \begin{split}
       A &= \eta a_R \left(\frac{(\Gamma-1)\mu u}{\kappa_B}\right)^4 \\
        B &= (1+\eta)\rho\hat{c}/c \\
        C &= E'_r - B\left(\frac{E_\mathrm{tot}}{\rho}-\frac{\mathbf{v}^2}{2}\right)\,,
    \end{split}
 \end{equation}
 where we have defined $\eta = \hat{c}\rho\kappa s \Delta t$. We solve this system iteratively by first updating the guess for $\mathbf{F}_r$ using Eq. \ref{Eq:ImplEq2}, as
 \begin{equation}\label{Eq:ImplFr}
     \mathbf{F}_r=\frac{1}{1+s\hat{c}\rho\chi\Delta t }\mathbf{F}_r'\,,
 \end{equation}
 after which we update $\mathbf{v}$ using Eq. \eqref{Eq:EFConsEqs}. This is done only once if constant opacities are used, in which case this expression satisfies Eq. \eqref{Eq:ImplEq2} exactly. We then compute the polynomial coefficients using Eq. \eqref{Eq:ImplABC} and apply Newton's method to update our guess of $\epsilon$, for which we neglect the derivatives of $A$, $B$ and $C$ with $\epsilon$. This yields a reasonable approximation to the derivative of the left-hand side of Eq. \eqref{Eq:ImplPol} as long as $\kappa$ and $\chi$ do not change abruptly with $\epsilon$ in the range spanned during the implicit step. This process is then repeated updating the opacities after every step if necessary until $\epsilon$ and $\mathbf{v}$ converge. Finally, $E$ is retrieved from $\epsilon$ and $\mathbf{v}$ using Eq. \eqref{Eq:GasEnergyDensity} and $E_r$ is computed using Eq. \eqref{Eq:EFConsEqs}.
    
  \section{Resolution study}\label{A:ResStudy}

   We tested if our Rad-HD simulations are affected by the employment of low resolutions per scale height in some regions of the disk, which could enhance the damping of forming radial structures. We considered for this the runs with the highest $f_\mathrm{dg}$, which are the most favorable ones for the potential development of shadowing structures. We repeated run \sftw{rhdg2\_c4} at the resolutions of $N_r\times N_\theta = 1024\times 400$ and $2048\times 800$, the last of which corresponds to $8-32$ cells in the $r$-direction and $24-120$ cells in the $\theta$-direction per scale height. The value $\hat{c}/c=10^{-4}$ was chosen for numerical efficiency since, as shown in Section \ref{SS:RSLA}, it is high enough not to alter the results.
    
\begin{figure*}[t]{
    \centering
    \includegraphics[width=\linewidth]{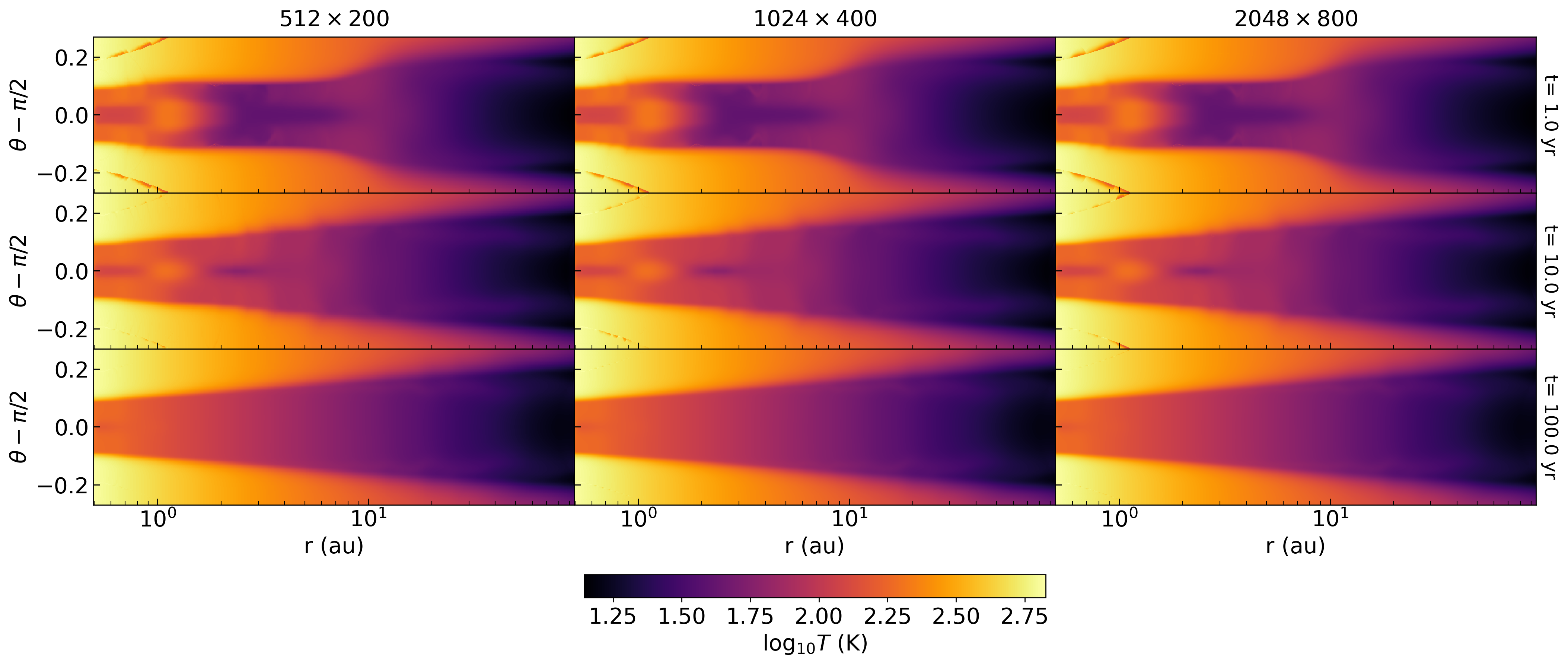}
    \caption{Temperature distributions as a function of time for run \sftw{rhdg2\_c4} (left column) and the corresponding higher-resolution runs (middle and right columns for $N_r\times N_\theta = 1024\times 400$ and $2048\times 800$, respectively).}
    \label{fig:rh_T2D_res}
    }

  \end{figure*}

  Figure \ref{fig:rh_T2D_res} shows a comparison of the temperature distributions in \sftw{rhdg2\_c4} and the corresponding higher-resolution runs at $t=1$, $10$, and $100$ yr. These distributions are almost indistinguishable at all times. In particular, the vertical extent of the thermally relaxed layers does not depend on the resolution, which shows that the vertical diffusion time is the same in all cases. This resolution independence indicates that numerical diffusion can only be of marginal relevance for the disk relaxation compared to the radiative and advection processes described in Section \ref{SS:RadHDSims}. This is also supported by the fact that radial structures formed through self-shadowing in the hydrostatic runs, occurring over typical lengthscales of a few au, are well resolved by our nominal resolution. As shown in Section \ref{SS:ResultingModels}, that resolution also suffices to reproduce the scale height variations in the $\theta$-direction, which are typically of order $H$.
  
\newpage

\begin{acknowledgements}
We thank Mario Flock for several helpful and insightful comments and discussions during this project. The research of J.D.M.F. and H.K. is supported by the German Science Foundation (DFG) under the priority program SPP 1992: “Exoplanet Diversity” under contract KL 1469/16-1/2. We thank our collaboration partners on this project in Kiel Sebastian Wolf and Anton Krieger, under contract WO 857/17-1/2, for providing us with the employed tabulated opacity coefficients. All numerical simulations were run on the ISAAC and VERA clusters of the MPIA and the COBRA cluster of the Max Planck Society, all of these hosted at the Max-Planck Computing and Data Facility in Garching (Germany). We also thank the anonymous referee for constructive comments that helped to improve the quality of this work.
\end{acknowledgements}
\bibliography{refs}

\end{document}